\documentclass[12pt]{article}
\usepackage{clipboard}%For copy past text in different parts of document
\usepackage{amsmath, amssymb,mathrsfs} %For using math
\usepackage{color}
\usepackage{enumitem}
\usepackage{graphicx}
\usepackage{grffile}
\usepackage[utf8]{inputenx}% For proper input encoding
\usepackage{mathptmx}
\usepackage{pdflscape}
\usepackage{soul}
\usepackage[euler-digits]{eulervm}% A pretty math font

\usepackage{abstract}
\usepackage[margin=1in]{geometry} %For controlling the dimensions. Very useful for Posters, etc
\usepackage{fancyhdr}
\usepackage{titlesec}
\titleformat*{\section}{\normalsize\bfseries}
\titleformat*{\subsection}{\small\bfseries}
\titleformat*{\subsubsection}{\small\bfseries}
\titleformat*{\paragraph}{\small\bfseries}
\titleformat*{\subparagraph}{\small\bfseries}
\titlespacing\section{0pt}{6pt plus 4pt minus 2pt}{0pt plus 2pt minus 2pt}
\titlespacing\subsection{0pt}{6pt plus 4pt minus 2pt}{0pt plus 2pt minus 2pt}
\titlespacing\subsubsection{0pt}{6pt plus 4pt minus 2pt}{0pt plus 2pt minus 2pt}
\titlespacing\paragraph{0pt}{6pt plus 4pt minus 2pt}{6pt plus 2pt minus 2pt}

\usepackage[authoryear]{natbib}	%Bibliography package. It handles citations and can easily change formats. This is done at the end of the document
\usepackage{multibib}
\newcites{App}{Appendix References}%  \citelatex, \nocitelatex, ...
\newcites{Ref}{References in Referee Replies}%  \citelatex, \nocitelatex, ...
\newcites{Main}{References}%  \citelatex, \nocitelatex, ...

\setcitestyle{aysep={}} 
\usepackage{accents}
\usepackage{adjustbox} %Handles resizing of tables, figures, etc. Based off of page and/or line width/height 
\usepackage{bbm}
\usepackage{bm}
\usepackage{booktabs}% For Pretty tables
\usepackage{caption} %\captionsetup[figure]{justification=raggedright,singlelinecheck=off}
\usepackage{chngpage} 
\usepackage{comment}
\usepackage{dcolumn}
\usepackage[capposition=top]{floatrow}
\usepackage{morefloats}
\usepackage{multicol}
\usepackage{multirow}
\usepackage{placeins}% To Create Float Barriers so the tables will stay in their sections
\usepackage[list=true]{subcaption} %For having multiple figures within the same one. Figure 1, with part (a) and (b) %\setcounter{lofdepth}{2}
\usepackage{threeparttable}% For Notes below table
\usepackage{rotating}% To Rotate Table
\usepackage{setspace} %Single, Double Space, etc 
\usepackage{siunitx} %This aligns tables by their decimal and handles processing of the numbers within the tables
\AtBeginDocument{
    \addtolength{\abovedisplayskip}{-1.0ex}
    \addtolength{\abovedisplayshortskip}{-1.0ex}
    \addtolength{\belowdisplayskip}{-1.0ex}
    \addtolength{\belowdisplayshortskip}{-1.0ex}
}
\DeclareCaptionLabelFormat{blank}{}
\DeclareCaptionFormat{myformat}{
    \begin{varwidth}{\linewidth}
        \centering
        #1#2#3%
    \end{varwidth}
}
\usepackage{silence}
\usepackage[unicode=true,pdfusetitle,bookmarks=true,bookmarksnumbered=false,bookmarksopen=false,breaklinks=true,pdfborder={0 0 1},backref=false,colorlinks=true,citecolor=black,allcolors=black,hyperfootnotes=true]{hyperref}
\usepackage[titletoc]{appendix}

\usepackage{xcolor}
\usepackage{xargs}
\renewcommand{\thefootnote}{\fnsymbol{footnote}}

\makeatletter
\AddToHook{env/tabular/begin}{\let\estinput=\@@input}
\makeatother
\DeclareUnicodeCharacter{00A0}{ }
\usepackage{ifthen}
\newboolean{stars_on}
\setboolean{stars_on}{false} % or {false}
\ifthenelse{\boolean{stars_on}}
{%
  \newcommand{\sym}[1]{\rlap{#1}}

}
{%
  \newcommand{\sym}[1]{\rlap{~}}

}
\def\yyy{%
  \bgroup\uccode`\~\expandafter`\string-%
  \uppercase{\egroup\edef~{\noexpand\text{\llap{\textendash}\relax}}}%
  \mathcode\expandafter`\string-"8000 }

\def\xxxl#1{%
\bgroup\uccode`\~\expandafter`\string#1%
\uppercase{\egroup\edef~{\noexpand\text{\noexpand\llap{\string#1}}}}%
\mathcode\expandafter`\string#1"8000 }

\def\xxxr#1{%
\bgroup\uccode`\~\expandafter`\string#1%
\uppercase{\egroup\edef~{\noexpand\text{\noexpand\rlap{\string#1}}}}%
\mathcode\expandafter`\string#1"8000 }

\catcode`_11
\protected\def \c__siunitx_minus_tl {$-$}
\catcode`_ 8 

\makeatletter 
\edef\originalbmathcode{%
    \noexpand\mathchardef\noexpand\@tempa\the\mathcode`\(\relax} %\)

\def\resetMathstrut@{%
  \setbox\z@\hbox{%
    \originalbmathcode

    \def\@tempb##1"##2##3{\the\textfont"##3\char"}%
    \expandafter\@tempb\meaning\@tempa \relax
  }%
  \ht\Mathstrutbox@\ht\z@ \dp\Mathstrutbox@\dp\z@
}
\makeatother %./tables.tex:25: LaTeX Error: Bad math environment delimiter. [\)]
\usepackage{xparse}

\ExplSyntaxOn
\DeclareExpandableDocumentCommand{\append}{mmm}
 {
  #1\prg_replicate:nn{#3}{#2}
 }
\ExplSyntaxOff

\newcommand{\noteDate}{November 18, 2026}

\newcommand{\pdslabel}{Agreement with the cleaned validation sample; scan-verified accuracy not assessed.}

\newcommand{\nPagesFrame}{1,036}
\newcommand{\nPagesBroad}{1,233}
\newcommand{\nRowsCleaned}{124,167}
\newcommand{\nRowsCleanedBroad}{145,064}
\newcommand{\precA}{94.11}
\newcommand{\recA}{94.13}
\newcommand{\precAstrict}{93.89}
\newcommand{\recAstrict}{93.91}
\newcommand{\nOutA}{124,195}
\newcommand{\nMatchedA}{116,883}
\newcommand{\nStrictA}{116,602}
\newcommand{\nSecondA}{281}
\newcommand{\precAbroad}{92.55}
\newcommand{\recAbroad}{93.96}
\newcommand{\precB}{96.72}
\newcommand{\recB}{97.45}
\newcommand{\precBstrict}{96.66}
\newcommand{\recBstrict}{97.39}
\newcommand{\nOutB}{125,101}
\newcommand{\nMatchedB}{120,998}
\newcommand{\nStrictB}{120,923}
\newcommand{\nSecondB}{75}
\newcommand{\precBbroad}{95.11}
\newcommand{\recBbroad}{97.44}
\newcommand{\precC}{98.04}
\newcommand{\recC}{98.28}
\newcommand{\precCstrict}{97.97}
\newcommand{\recCstrict}{98.21}
\newcommand{\nOutC}{124,477}
\newcommand{\nMatchedC}{122,032}
\newcommand{\nStrictC}{121,949}
\newcommand{\nSecondC}{83}
\newcommand{\precCbroad}{96.43}
\newcommand{\recCbroad}{98.32}
\newcommand{\precD}{96.06}
\newcommand{\recD}{95.09}
\newcommand{\precDstrict}{95.99}
\newcommand{\recDstrict}{95.02}
\newcommand{\nOutD}{122,915}
\newcommand{\nMatchedD}{118,070}
\newcommand{\nStrictD}{117,987}
\newcommand{\nSecondD}{83}
\newcommand{\precDbroad}{94.44}
\newcommand{\recDbroad}{95.22}

\newcommand{\nPdsNoKeyC}{1,569}
\newcommand{\nPdsAmbiguousC}{107}
\newcommand{\nPdsMissingKeyC}{459}
\newcommand{\nOutNoKeyC}{2,327}
\newcommand{\nOutAmbiguousC}{118}

\newcommand{\precStageLoc}{95.44}
\newcommand{\precStageLocxii}{93.77}
\newcommand{\recStageLoc}{91.64}
\newcommand{\recStageLocxii}{90.34}
\newcommand{\precStageBio}{95.64}
\newcommand{\precStageBioxii}{96.46}
\newcommand{\recStageBio}{92.68}
\newcommand{\recStageBioxii}{92.10}
\newcommand{\precStageMatchNoR}{94.76}
\newcommand{\precStageMatchNoRxii}{95.09}
\newcommand{\recStageMatchNoR}{92.77}
\newcommand{\recStageMatchNoRxii}{92.13}
\newcommand{\nRFourLinksAdded}{1,691}
\newcommand{\nRFourLinksAddedxii}{2,824}
\newcommand{\nRFourRows}{2,958}
\newcommand{\nRFourRowsxii}{3,654}
\newcommand{\nRFourRowsFrame}{2,632}
\newcommand{\nRFourRowsFramexii}{3,227}
\newcommand{\fieldNameA}{95.06 / 94.48}

\newcommand{\cntNameA}{116,883 / 122,953 / 123,708}
\newcommand{\fieldNameB}{96.72 / 97.81}

\newcommand{\cntNameB}{120,998 / 125,101 / 123,708}
\newcommand{\fieldNameC}{98.04 / 98.65}

\newcommand{\cntNameC}{122,032 / 124,477 / 123,708}
\newcommand{\fieldNameD}{96.06 / 95.44}

\newcommand{\cntNameD}{118,070 / 122,915 / 123,708}
\newcommand{\cntNameE}{115,136 / 121,295 / 123,708}
\newcommand{\cntNameAxii}{116,817 / 121,356 / 122,188}
\newcommand{\cntNameBxii}{120,462 / 123,496 / 122,188}
\newcommand{\cntNameCxii}{121,010 / 123,829 / 122,188}
\newcommand{\cntNameDxii}{118,613 / 121,370 / 122,188}
\newcommand{\cntNameExii}{115,780 / 119,991 / 122,188}
\newcommand{\fieldBirthYearA}{95.00 / 96.13}

\newcommand{\cntBirthYearA}{105,239 / 110,774 / 109,480}
\newcommand{\fieldBirthYearB}{97.33 / 97.57}

\newcommand{\cntBirthYearB}{106,823 / 109,749 / 109,480}
\newcommand{\fieldBirthYearC}{98.00 / 98.30}

\newcommand{\cntBirthYearC}{107,624 / 109,825 / 109,480}
\newcommand{\fieldBirthYearD}{96.29 / 95.03}

\newcommand{\cntBirthYearD}{104,037 / 108,043 / 109,480}
\newcommand{\cntBirthYearE}{101,355 / 106,873 / 109,480}
\newcommand{\cntBirthYearAxii}{87,462 / 90,301 / 90,208}
\newcommand{\cntBirthYearBxii}{88,130 / 90,014 / 90,208}
\newcommand{\cntBirthYearCxii}{88,557 / 90,308 / 90,208}
\newcommand{\cntBirthYearDxii}{86,459 / 88,212 / 90,208}
\newcommand{\cntBirthYearExii}{84,086 / 87,829 / 90,208}
\newcommand{\fieldSchoolA}{94.65 / 89.72}

\newcommand{\cntSchoolA}{108,682 / 114,831 / 121,137}
\newcommand{\fieldSchoolB}{96.94 / 96.74}

\newcommand{\cntSchoolB}{117,191 / 120,890 / 121,137}
\newcommand{\fieldSchoolC}{97.45 / 98.03}

\newcommand{\cntSchoolC}{118,746 / 121,851 / 121,137}
\newcommand{\fieldSchoolD}{94.48 / 92.49}

\newcommand{\fieldSchoolRD}{92.49}
\newcommand{\cntSchoolD}{112,038 / 118,578 / 121,137}
\newcommand{\cntSchoolE}{109,605 / 117,862 / 121,137}
\newcommand{\cntSchoolAxii}{102,980 / 109,737 / 119,787}
\newcommand{\cntSchoolBxii}{114,134 / 119,140 / 119,787}
\newcommand{\cntSchoolCxii}{115,109 / 119,902 / 119,787}
\newcommand{\cntSchoolDxii}{109,317 / 114,258 / 119,787}
\newcommand{\cntSchoolExii}{104,060 / 110,891 / 119,787}
\newcommand{\fieldGradYearA}{94.03 / 94.85}

\newcommand{\cntGradYearA}{112,852 / 120,019 / 118,984}
\newcommand{\fieldGradYearB}{95.44 / 96.86}

\newcommand{\cntGradYearB}{115,252 / 120,764 / 118,984}
\newcommand{\fieldGradYearC}{95.88 / 98.24}

\newcommand{\cntGradYearC}{116,894 / 121,920 / 118,984}
\newcommand{\fieldGradYearD}{95.90 / 94.44}

\newcommand{\cntGradYearD}{112,374 / 117,178 / 118,984}
\newcommand{\cntGradYearE}{109,326 / 115,622 / 118,984}
\newcommand{\cntGradYearAxii}{112,071 / 117,404 / 117,731}
\newcommand{\cntGradYearBxii}{114,101 / 119,124 / 117,731}
\newcommand{\cntGradYearCxii}{115,193 / 120,326 / 117,731}
\newcommand{\cntGradYearDxii}{112,291 / 115,385 / 117,731}
\newcommand{\cntGradYearExii}{108,787 / 113,771 / 117,731}
\newcommand{\fieldLicenceYearA}{93.06 / 90.66}

\newcommand{\cntLicenceYearA}{93,603 / 100,580 / 103,249}
\newcommand{\fieldLicenceYearB}{96.78 / 97.29}

\newcommand{\cntLicenceYearB}{100,454 / 103,791 / 103,249}
\newcommand{\fieldLicenceYearC}{96.85 / 98.21}

\newcommand{\cntLicenceYearC}{101,401 / 104,701 / 103,249}
\newcommand{\fieldLicenceYearD}{95.73 / 94.53}

\newcommand{\cntLicenceYearD}{97,604 / 101,961 / 103,249}
\newcommand{\cntLicenceYearE}{92,869 / 99,265 / 103,249}
\newcommand{\cntLicenceYearAxii}{85,040 / 90,908 / 93,693}
\newcommand{\cntLicenceYearBxii}{91,205 / 94,036 / 93,693}
\newcommand{\cntLicenceYearCxii}{91,669 / 95,786 / 93,693}
\newcommand{\cntLicenceYearDxii}{89,678 / 92,491 / 93,693}
\newcommand{\cntLicenceYearExii}{85,107 / 91,852 / 93,693}
\newcommand{\fieldCityA}{88.98 / 85.68}

\newcommand{\cntCityA}{106,196 / 119,344 / 123,948}
\newcommand{\fieldCityB}{95.25 / 96.14}

\newcommand{\cntCityB}{119,158 / 125,101 / 123,948}
\newcommand{\fieldCityC}{96.58 / 96.95}

\newcommand{\cntCityC}{120,164 / 124,424 / 123,948}
\newcommand{\fieldCityD}{91.76 / 91.00}

\newcommand{\cntCityD}{112,788 / 122,915 / 123,948}
\newcommand{\cntCityE}{111,167 / 121,295 / 123,948}
\newcommand{\cntCityAxii}{111,237 / 119,280 / 123,713}
\newcommand{\cntCityBxii}{118,282 / 123,371 / 123,713}
\newcommand{\cntCityCxii}{119,470 / 123,768 / 123,713}
\newcommand{\cntCityDxii}{115,220 / 121,370 / 123,713}
\newcommand{\cntCityExii}{111,960 / 119,991 / 123,713}
\newcommand{\fieldCountyA}{83.36 / 80.28}

\newcommand{\cntCountyA}{99,482 / 119,344 / 123,925}
\newcommand{\fieldCountyB}{95.14 / 89.78}

\newcommand{\cntCountyB}{111,256 / 116,944 / 123,925}
\newcommand{\fieldCountyC}{96.65 / 96.97}

\newcommand{\cntCountyC}{120,170 / 124,334 / 123,925}
\newcommand{\fieldCountyD}{86.31 / 85.61}

\newcommand{\cntCountyD}{106,087 / 122,915 / 123,925}
\newcommand{\cntCountyE}{105,855 / 121,241 / 123,925}
\newcommand{\cntCountyAxii}{105,991 / 119,280 / 123,725}
\newcommand{\cntCountyBxii}{106,655 / 112,414 / 123,725}
\newcommand{\cntCountyCxii}{119,630 / 123,702 / 123,725}
\newcommand{\cntCountyDxii}{111,035 / 121,370 / 123,725}
\newcommand{\cntCountyExii}{106,128 / 119,991 / 123,725}
\newcommand{\fieldStateA}{94.24 / 90.58}

\newcommand{\cntStateA}{112,472 / 119,344 / 124,165}
\newcommand{\fieldStateB}{96.72 / 97.45}

\newcommand{\cntStateB}{120,998 / 125,101 / 124,165}
\newcommand{\fieldStateC}{98.06 / 98.27}

\newcommand{\cntStateC}{122,013 / 124,424 / 124,165}
\newcommand{\fieldStateD}{96.06 / 95.08}

\newcommand{\cntStateD}{118,053 / 122,897 / 124,165}
\newcommand{\cntStateE}{114,488 / 121,146 / 124,165}
\newcommand{\cntStateAxii}{111,837 / 118,200 / 123,725}
\newcommand{\cntStateBxii}{120,461 / 123,496 / 123,725}
\newcommand{\cntStateCxii}{120,982 / 123,769 / 123,725}
\newcommand{\cntStateDxii}{117,799 / 121,129 / 123,725}
\newcommand{\cntStateExii}{115,294 / 119,489 / 123,725}

\newcommand{\nLinkedRowsVC}{122,032}
\newcommand{\nLinkedRowsVCxii}{121,010}
\newcommand{\nLinkedPagesVC}{1,034}
\newcommand{\nLinkedPagesVCxii}{903}
\newcommand{\disBirthYearGeminiRight}{381}
\newcommand{\disBirthYearGeminiRightxii}{231}
\newcommand{\disBirthYearParserRight}{1,893}
\newcommand{\disBirthYearParserRightxii}{2,213}
\newcommand{\disBirthYearBothAgree}{105,353}
\newcommand{\disBirthYearBothAgreexii}{86,130}
\newcommand{\disBirthYearNeither}{64}
\newcommand{\disBirthYearNeitherxii}{77}
\newcommand{\disBirthYearParserRightRowAbsent}{1,798}

\newcommand{\disGradYearGeminiRight}{1,228}
\newcommand{\disGradYearGeminiRightxii}{847}
\newcommand{\disGradYearParserRight}{3,031}
\newcommand{\disGradYearParserRightxii}{3,703}
\newcommand{\disGradYearBothAgree}{112,659}
\newcommand{\disGradYearBothAgreexii}{110,682}
\newcommand{\disGradYearNeither}{118}
\newcommand{\disGradYearNeitherxii}{192}
\newcommand{\disGradYearParserRightRowAbsent}{1,958}

\newcommand{\disLicenceYearGeminiRight}{534}
\newcommand{\disLicenceYearGeminiRightxii}{269}
\newcommand{\disLicenceYearParserRight}{5,676}
\newcommand{\disLicenceYearParserRightxii}{7,911}
\newcommand{\disLicenceYearBothAgree}{95,196}
\newcommand{\disLicenceYearBothAgreexii}{83,524}
\newcommand{\disLicenceYearNeither}{120}
\newcommand{\disLicenceYearNeitherxii}{160}
\newcommand{\disLicenceYearParserRightRowAbsent}{1,594}

\newcommand{\disSchoolGeminiRight}{1,140}
\newcommand{\disSchoolGeminiRightxii}{1,067}
\newcommand{\disSchoolParserRight}{7,106}
\newcommand{\disSchoolParserRightxii}{11,530}
\newcommand{\disSchoolBothAgree}{108,471}
\newcommand{\disSchoolBothAgreexii}{101,130}
\newcommand{\disSchoolNeither}{318}
\newcommand{\disSchoolNeitherxii}{1,718}
\newcommand{\disSchoolParserRightRowAbsent}{1,955}

\newcommand{\cleanSchoolAsRead}{97.45 / 98.03}
\newcommand{\cleanSchoolAsReadxii}{96.00 / 96.09}
\newcommand{\cleanSchoolCleaned}{97.48 / 98.11}
\newcommand{\cleanSchoolCleanedxii}{97.47 / 97.59}
\newcommand{\cleanBirthYearAsRead}{98.00 / 98.30}
\newcommand{\cleanBirthYearAsReadxii}{98.06 / 98.17}
\newcommand{\cleanBirthYearCleaned}{98.01 / 98.30}
\newcommand{\cleanBirthYearCleanedxii}{98.07 / 98.16}
\newcommand{\cleanCityAsRead}{96.58 / 96.95}
\newcommand{\cleanCityAsReadxii}{96.53 / 96.57}
\newcommand{\cleanCityCleaned}{96.58 / 96.95}
\newcommand{\cleanCityCleanedxii}{96.53 / 96.57}
\newcommand{\cleanCountyAsRead}{90.90 / 90.96}
\newcommand{\cleanCountyAsReadxii}{91.34 / 91.26}
\newcommand{\cleanCountyCleaned}{96.65 / 96.97}
\newcommand{\cleanCountyCleanedxii}{96.71 / 96.69}

\newcommand{\nDeliveryRowsVzero}{148,040}

\newcommand{\releaseVersion}{1.0.0}
\newcommand{\zenodoDOI}{\releaseDOI} % alias: generated from docs/releases/zenodo-1.0.0.yaml

\newcommand{\geminiModelName}{gemini-2.0-flash}
\newcommand{\geminiShutdownDate}{1 June 2026}

\newcommand{\nStates}{48}

\newcommand{\nChrisingerEntries}{41,828}
\newcommand{\nChrisingerLabelled}{632}
\newcommand{\nChrisingerBlackSchool}{113}
\newcommand{\smithEntriesKept}{70 to 90}
\newcommand{\smithEntriesPerfect}{92}

\newcommand{\nPagesFramexii}{903}
\newcommand{\nRowsCleanedxii}{123,726}
\newcommand{\nRowsCleanedBroadxii}{137,501}
\newcommand{\nPagesBroadxii}{1,006}
\newcommand{\nOutAxii}{123,108}
\newcommand{\nStrictAxii}{116,765}
\newcommand{\nSecondAxii}{52}
\newcommand{\nMatchedAxii}{116,817}
\newcommand{\precAxii}{94.89}
\newcommand{\recAxii}{94.42}
\newcommand{\precAstrictxii}{94.85}
\newcommand{\recAstrictxii}{94.37}
\newcommand{\precAbroadxii}{94.76}
\newcommand{\recAbroadxii}{94.21}
\newcommand{\nOutBxii}{123,496}
\newcommand{\nStrictBxii}{120,420}
\newcommand{\nSecondBxii}{42}
\newcommand{\nMatchedBxii}{120,462}
\newcommand{\precBxii}{97.54}
\newcommand{\recBxii}{97.36}
\newcommand{\precBstrictxii}{97.51}
\newcommand{\recBstrictxii}{97.33}
\newcommand{\precBbroadxii}{97.48}
\newcommand{\recBbroadxii}{97.27}
\newcommand{\nOutCxii}{123,829}
\newcommand{\nStrictCxii}{120,965}
\newcommand{\nSecondCxii}{45}
\newcommand{\nMatchedCxii}{121,010}
\newcommand{\precCxii}{97.72}
\newcommand{\recCxii}{97.80}
\newcommand{\precCstrictxii}{97.69}
\newcommand{\recCstrictxii}{97.77}
\newcommand{\precCbroadxii}{97.65}
\newcommand{\recCbroadxii}{97.72}
\newcommand{\nOutDxii}{121,370}
\newcommand{\nStrictDxii}{118,571}
\newcommand{\nSecondDxii}{42}
\newcommand{\nMatchedDxii}{118,613}
\newcommand{\precDxii}{97.73}
\newcommand{\recDxii}{95.87}
\newcommand{\precDstrictxii}{97.69}
\newcommand{\recDstrictxii}{95.83}
\newcommand{\precDbroadxii}{97.57}
\newcommand{\recDbroadxii}{95.60}
\newcommand{\fieldNameAxii}{96.26 / 95.60}
\newcommand{\fieldNameBxii}{97.54 / 98.59}
\newcommand{\fieldNameCxii}{97.72 / 99.04}
\newcommand{\fieldNameDxii}{97.73 / 97.07}
\newcommand{\fieldBirthYearAxii}{96.86 / 96.96}
\newcommand{\fieldBirthYearBxii}{97.91 / 97.70}
\newcommand{\fieldBirthYearCxii}{98.06 / 98.17}
\newcommand{\fieldBirthYearDxii}{98.01 / 95.84}
\newcommand{\fieldSchoolAxii}{93.84 / 85.97}
\newcommand{\fieldSchoolBxii}{95.80 / 95.28}
\newcommand{\fieldSchoolCxii}{96.00 / 96.09}
\newcommand{\fieldSchoolDxii}{95.68 / 91.26}
\newcommand{\fieldSchoolRDxii}{91.26}
\newcommand{\fieldGradYearAxii}{95.46 / 95.19}
\newcommand{\fieldGradYearBxii}{95.78 / 96.92}
\newcommand{\fieldGradYearCxii}{95.73 / 97.84}
\newcommand{\fieldGradYearDxii}{97.32 / 95.38}
\newcommand{\fieldLicenceYearAxii}{93.55 / 90.76}
\newcommand{\fieldLicenceYearBxii}{96.99 / 97.34}
\newcommand{\fieldLicenceYearCxii}{95.70 / 97.84}
\newcommand{\fieldLicenceYearDxii}{96.96 / 95.71}
\newcommand{\fieldStateAxii}{94.62 / 90.39}
\newcommand{\fieldStateBxii}{97.54 / 97.36}
\newcommand{\fieldStateCxii}{97.75 / 97.78}
\newcommand{\fieldStateDxii}{97.25 / 95.21}
\newcommand{\fieldCityAxii}{93.26 / 89.92}
\newcommand{\fieldCityBxii}{95.88 / 95.61}
\newcommand{\fieldCityCxii}{96.53 / 96.57}
\newcommand{\fieldCityDxii}{94.93 / 93.13}
\newcommand{\fieldCountyAxii}{88.86 / 85.67}
\newcommand{\fieldCountyBxii}{94.88 / 86.20}
\newcommand{\fieldCountyCxii}{96.71 / 96.69}
\newcommand{\fieldCountyDxii}{91.48 / 89.74}
\newcommand{\nOutE}{121,295}
\newcommand{\nMatchedE}{115,136}
\newcommand{\precE}{94.92}
\newcommand{\recE}{92.73}
\newcommand{\fieldNameE}{94.92 / 93.07}
\newcommand{\fieldBirthYearE}{94.84 / 92.58}
\newcommand{\fieldSchoolE}{92.99 / 90.48}
\newcommand{\fieldGradYearE}{94.55 / 91.88}
\newcommand{\fieldLicenceYearE}{93.56 / 89.95}
\newcommand{\fieldStateE}{94.50 / 92.21}
\newcommand{\fieldCityE}{91.65 / 89.69}
\newcommand{\fieldCountyE}{87.31 / 85.42}
\newcommand{\nOutExii}{119,991}
\newcommand{\nMatchedExii}{115,780}
\newcommand{\precExii}{96.49}
\newcommand{\recExii}{93.58}
\newcommand{\fieldNameExii}{96.49 / 94.76}
\newcommand{\fieldBirthYearExii}{95.74 / 93.21}
\newcommand{\fieldSchoolExii}{93.84 / 86.87}
\newcommand{\fieldGradYearExii}{95.62 / 92.40}
\newcommand{\fieldLicenceYearExii}{92.66 / 90.84}
\newcommand{\fieldStateExii}{96.49 / 93.19}
\newcommand{\fieldCityExii}{93.31 / 90.50}
\newcommand{\fieldCountyExii}{88.45 / 85.78}
\newcommand{\nDeliveryRowsxii}{138,800}
\newcommand{\nEvalRowsDVzero}{162,215}
\newcommand{\nOutputRowsDVzero}{162,947}
\newcommand{\nDupPageRowsDVzero}{732}

\newcommand{\nPagesRobAll}{1,233}
\newcommand{\nRowsRobAll}{145,064}
\newcommand{\precRobAllA}{92.55}
\newcommand{\recRobAllA}{93.96}
\newcommand{\precRobAllB}{95.11}
\newcommand{\recRobAllB}{97.44}
\newcommand{\precRobAllC}{96.43}
\newcommand{\recRobAllC}{98.32}
\newcommand{\precRobAllD}{94.44}
\newcommand{\recRobAllD}{95.22}
\newcommand{\precRobAllE}{93.35}
\newcommand{\recRobAllE}{92.74}
\newcommand{\nPagesRobIncomplete}{197}
\newcommand{\nRowsRobIncomplete}{20,897}
\newcommand{\precRobIncompleteA}{84.11}
\newcommand{\recRobIncompleteA}{92.90}
\newcommand{\precRobIncompleteB}{86.54}
\newcommand{\recRobIncompleteB}{97.42}
\newcommand{\precRobIncompleteC}{87.89}
\newcommand{\recRobIncompleteC}{98.53}
\newcommand{\precRobIncompleteD}{85.90}
\newcommand{\recRobIncompleteD}{96.00}
\newcommand{\precRobIncompleteE}{85.02}
\newcommand{\recRobIncompleteE}{92.85}
\newcommand{\nPagesRobOld}{902}
\newcommand{\nRowsRobOld}{107,565}
\newcommand{\precRobOldA}{94.33}
\newcommand{\recRobOldA}{94.29}
\newcommand{\precRobOldB}{96.59}
\newcommand{\recRobOldB}{97.46}
\newcommand{\precRobOldC}{98.06}
\newcommand{\recRobOldC}{98.31}
\newcommand{\precRobOldD}{96.09}
\newcommand{\recRobOldD}{95.32}
\newcommand{\precRobOldE}{94.87}
\newcommand{\recRobOldE}{92.67}
\newcommand{\nPagesRobExposed}{134}
\newcommand{\nRowsRobExposed}{16,602}
\newcommand{\precRobExposedA}{92.74}
\newcommand{\recRobExposedA}{93.12}
\newcommand{\precRobExposedB}{97.59}
\newcommand{\recRobExposedB}{97.39}
\newcommand{\precRobExposedC}{97.87}
\newcommand{\recRobExposedC}{98.10}
\newcommand{\precRobExposedD}{95.84}
\newcommand{\recRobExposedD}{93.57}
\newcommand{\precRobExposedE}{95.28}
\newcommand{\recRobExposedE}{93.10}
\newcommand{\nPagesExposedDesign}{16}

\newcommand{\nPagesExposedTuning}{116}

\newcommand{\nPagesExposedDiagnostic}{2}

\newcommand{\nPagesUncHigh}{42}
\newcommand{\nPagesUncLow}{994}
\newcommand{\nRowsUncHigh}{5,085}
\newcommand{\nRowsEntUncUncertain}{1,627}
\newcommand{\nRowsEntUncCertain}{122,540}
\newcommand{\recEntUncUncertainA}{70.31}
\newcommand{\recEntUncUncertainB}{84.45}
\newcommand{\recEntUncUncertainC}{86.17}
\newcommand{\recEntUncUncertainD}{83.41}
\newcommand{\recEntUncUncertainE}{82.42}
\newcommand{\recEntUncCertainA}{94.45}
\newcommand{\recEntUncCertainB}{97.62}
\newcommand{\recEntUncCertainC}{98.44}
\newcommand{\recEntUncCertainD}{95.24}
\newcommand{\recEntUncCertainE}{92.86}
\newcommand{\nRowsUncLow}{119,082}
\newcommand{\precUncHighA}{92.18}
\newcommand{\recUncHighA}{90.84}
\newcommand{\precUncHighB}{93.34}
\newcommand{\recUncHighB}{94.75}
\newcommand{\precUncHighC}{94.03}
\newcommand{\recUncHighC}{95.36}
\newcommand{\precUncHighD}{92.43}
\newcommand{\recUncHighD}{93.35}
\newcommand{\precUncHighE}{92.05}
\newcommand{\recUncHighE}{91.72}
\newcommand{\precUncLowA}{94.19}
\newcommand{\recUncLowA}{94.27}
\newcommand{\precUncLowB}{96.87}
\newcommand{\recUncLowB}{97.56}
\newcommand{\precUncLowC}{98.21}
\newcommand{\recUncLowC}{98.41}
\newcommand{\precUncLowD}{96.22}
\newcommand{\recUncLowD}{95.16}
\newcommand{\precUncLowE}{95.05}
\newcommand{\recUncLowE}{92.77}
\newcommand{\nPagesRobAllxii}{1,006}
\newcommand{\nRowsRobAllxii}{137,501}
\newcommand{\precRobAllAxii}{94.76}
\newcommand{\recRobAllAxii}{94.21}
\newcommand{\precRobAllBxii}{97.48}
\newcommand{\recRobAllBxii}{97.27}
\newcommand{\precRobAllCxii}{97.65}
\newcommand{\recRobAllCxii}{97.72}
\newcommand{\precRobAllDxii}{97.57}
\newcommand{\recRobAllDxii}{95.60}
\newcommand{\precRobAllExii}{96.42}
\newcommand{\recRobAllExii}{93.54}
\newcommand{\nPagesRobIncompletexii}{103}
\newcommand{\nRowsRobIncompletexii}{13,775}
\newcommand{\precRobIncompleteAxii}{93.61}
\newcommand{\recRobIncompleteAxii}{92.40}
\newcommand{\precRobIncompleteBxii}{96.89}
\newcommand{\recRobIncompleteBxii}{96.48}
\newcommand{\precRobIncompleteCxii}{97.03}
\newcommand{\recRobIncompleteCxii}{96.93}
\newcommand{\precRobIncompleteDxii}{96.14}
\newcommand{\recRobIncompleteDxii}{93.15}
\newcommand{\precRobIncompleteExii}{95.77}
\newcommand{\recRobIncompleteExii}{93.21}
\newcommand{\nPagesRobOldxii}{794}
\newcommand{\nRowsRobOldxii}{108,531}
\newcommand{\precRobOldAxii}{95.01}
\newcommand{\recRobOldAxii}{94.64}
\newcommand{\precRobOldBxii}{97.62}
\newcommand{\recRobOldBxii}{97.46}
\newcommand{\precRobOldCxii}{97.80}
\newcommand{\recRobOldCxii}{97.89}
\newcommand{\precRobOldDxii}{97.78}
\newcommand{\recRobOldDxii}{95.94}
\newcommand{\precRobOldExii}{96.55}
\newcommand{\recRobOldExii}{93.64}
\newcommand{\nPagesRobExposedxii}{109}
\newcommand{\nRowsRobExposedxii}{15,195}
\newcommand{\precRobExposedAxii}{94.05}
\newcommand{\recRobExposedAxii}{92.83}
\newcommand{\precRobExposedBxii}{96.96}
\newcommand{\recRobExposedBxii}{96.66}
\newcommand{\precRobExposedCxii}{97.20}
\newcommand{\recRobExposedCxii}{97.20}
\newcommand{\precRobExposedDxii}{97.36}
\newcommand{\recRobExposedDxii}{95.31}
\newcommand{\precRobExposedExii}{96.07}
\newcommand{\recRobExposedExii}{93.16}
\newcommand{\nPagesExposedDesignxii}{16}

\newcommand{\nPagesExposedTuningxii}{93}

\newcommand{\nPagesExposedDiagnosticxii}{0}

\newcommand{\nPagesUncHighxii}{61}
\newcommand{\nPagesUncLowxii}{842}
\newcommand{\nRowsUncHighxii}{7,884}
\newcommand{\nRowsEntUncUncertainxii}{2,271}
\newcommand{\nRowsEntUncCertainxii}{121,455}
\newcommand{\recEntUncUncertainAxii}{45.13}
\newcommand{\recEntUncUncertainBxii}{46.81}
\newcommand{\recEntUncUncertainCxii}{47.07}
\newcommand{\recEntUncUncertainDxii}{46.15}
\newcommand{\recEntUncUncertainExii}{44.65}
\newcommand{\recEntUncCertainAxii}{95.34}
\newcommand{\recEntUncCertainBxii}{98.31}
\newcommand{\recEntUncCertainCxii}{98.75}
\newcommand{\recEntUncCertainDxii}{96.80}
\newcommand{\recEntUncCertainExii}{94.49}
\newcommand{\nRowsUncLowxii}{115,842}
\newcommand{\precUncHighAxii}{81.34}
\newcommand{\recUncHighAxii}{80.91}
\newcommand{\precUncHighBxii}{83.58}
\newcommand{\recUncHighBxii}{83.43}
\newcommand{\precUncHighCxii}{83.77}
\newcommand{\recUncHighCxii}{83.83}
\newcommand{\precUncHighDxii}{83.97}
\newcommand{\recUncHighDxii}{82.97}
\newcommand{\precUncHighExii}{82.30}
\newcommand{\recUncHighExii}{79.60}
\newcommand{\precUncLowAxii}{95.81}
\newcommand{\recUncLowAxii}{95.34}
\newcommand{\precUncLowBxii}{98.49}
\newcommand{\recUncLowBxii}{98.31}
\newcommand{\precUncLowCxii}{98.67}
\newcommand{\recUncLowCxii}{98.76}
\newcommand{\precUncLowDxii}{98.67}
\newcommand{\recUncLowDxii}{96.75}
\newcommand{\precUncLowExii}{97.45}
\newcommand{\recUncLowExii}{94.53}
\newcommand{\rangeFieldsBC}{95 to 98} % derived: round(min)-round(max) of P and R, four biographical fields, methods B and C, 1921
\newcommand{\rangeFieldsA}{90 to 96} % derived: round(min)-round(max) of P and R, four biographical fields, method A, 1921
\newcommand{\rangeFieldsD}{92 to 96} % derived: round(min)-round(max) of P and R, four biographical fields, method D, 1921
\newcommand{\rangeCityBC}{95 to 97} % derived: round(min)-round(max) of P and R, city, methods B and C, 1921
\newcommand{\rangeCityBCxii}{96 to 97} % derived: round(min)-round(max) of P and R, city, methods B and C, 1912
\newcommand{\rangeCityA}{86 to 89} % derived: round(min)-round(max) of P and R, city, method A, 1921
\newcommand{\rangeCityAxii}{90 to 93} % derived: round(min)-round(max) of P and R, city, method A, 1912
\newcommand{\rangeStageRec}{92 to 93} % derived: round(min)-round(max) of recall of the location and biography passes alone, 1921
\newcommand{\rangeStagePrec}{95 to 96} % derived: round(min)-round(max) of precision of the location and biography passes alone, 1921
\newcommand{\recGapBA}{3.31}

\newcommand{\recGapBD}{2.36}

\newcommand{\precGapBD}{0.66}

\newcommand{\releaseDOI}{10.5281/zenodo.23945788}
\IfFileExists{numbers.tex}{}{\errmessage{Missing reporting input: numbers.tex}\csname @@end\endcsname}
\ifnum\pdfstrcmp{\pdfmdfivesum file {numbers.tex}}{A94A7CFFF2AF908369BA8E64946EF57E}=0\else
\errmessage{Stale reporting input: numbers.tex. Regenerate and export from analysis}\csname @@end\endcsname\fi
\IfFileExists{field-precision-recall.pdf}{}{\errmessage{Missing reporting input: field-precision-recall.pdf}\csname @@end\endcsname}
\ifnum\pdfstrcmp{\pdfmdfivesum file {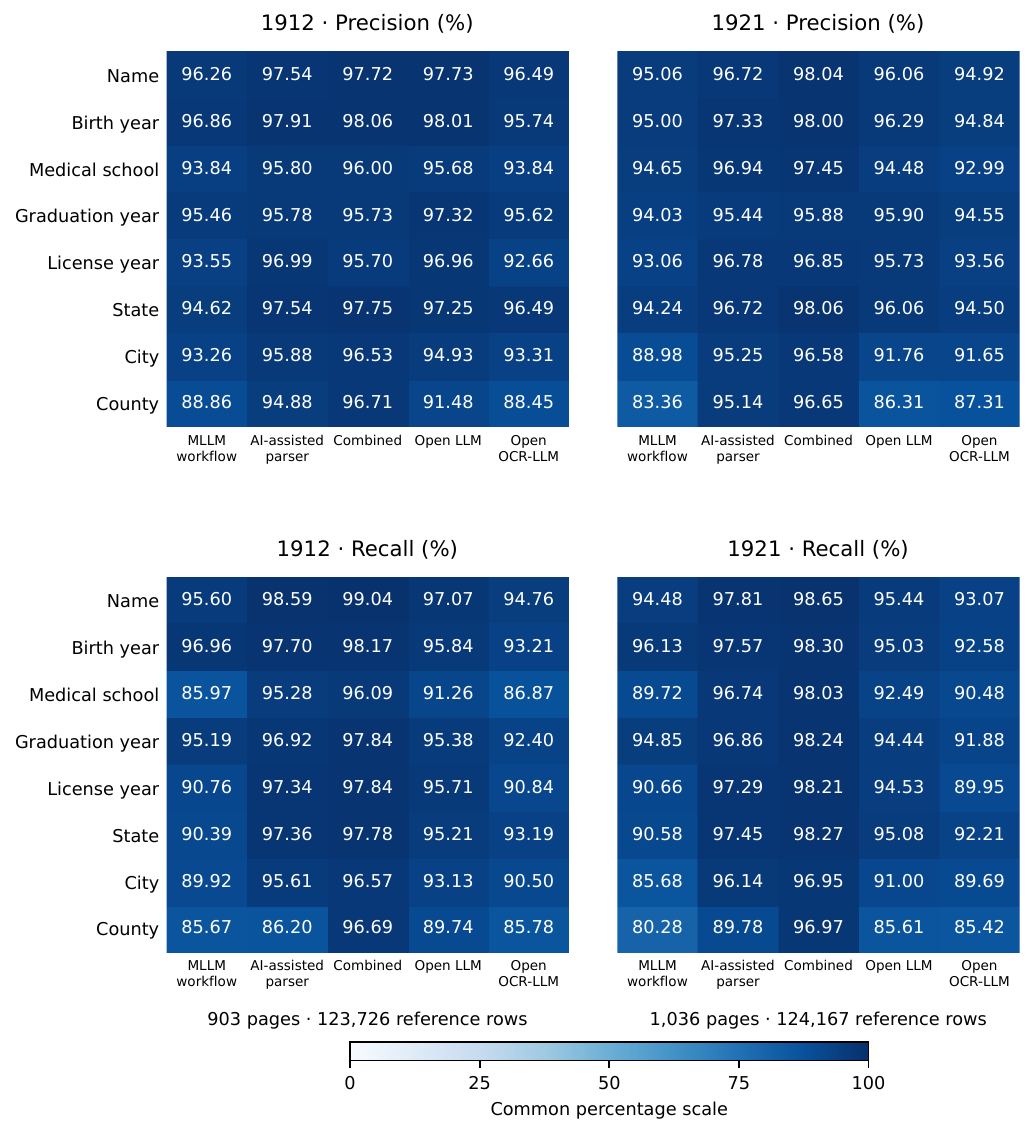}}{CBDEF759393FAFB97147DA3A72F503C5}=0\else
\errmessage{Stale reporting input: field-precision-recall.pdf. Regenerate and export from analysis}\csname @@end\endcsname\fi

\newcommand{\costManualPagesxii}{1,007}
\newcommand{\costOcrPagesxii}{1,388}
\newcommand{\costManualEurxii}{14,802.90}
\newcommand{\costGeminiDkkxii}{136.99}
\newcommand{\costAzureUsdxii}{13.88}
\newcommand{\costApiEurxii}{30.10}
\newcommand{\costApiPageEurxii}{0.03}

\newcommand{\costApiRatioxii}{492}
\newcommand{\costManualPages}{1,235}
\newcommand{\costOcrPages}{1,480}
\newcommand{\costManualEur}{18,154.50}
\newcommand{\costGeminiDkk}{241.71}
\newcommand{\costAzureUsd}{14.80}
\newcommand{\costApiEur}{44.90}
\newcommand{\costApiPageEur}{0.04}

\newcommand{\costApiRatio}{404}
\newcommand{\costManualPagesBoth}{2,242}

\newcommand{\costManualEurBoth}{32,957.40}
\newcommand{\costGeminiDkkBoth}{378.70}
\newcommand{\costAzureUsdBoth}{28.68}
\newcommand{\costApiEurBoth}{75.01}
\newcommand{\costApiPageEurBoth}{0.03}

\newcommand{\costApiRatioBoth}{439}

\newcommand{\costPdsPageEur}{14.70}
\newcommand{\costAzureThousandUsd}{10}

\newcommand{\costApiRatioRounded}{440}
\newcommand{\costGeminiSharedExcludedDkkxii}{114.12}
\newcommand{\costGeminiSharedExcludedDkk}{185.45}
\newcommand{\costFxDkkEur}{7.4687}
\newcommand{\costFxUsdEur}{1.1801}

\newcommand{\recGapDE}{\fpeval{round(\recD-\recE,1)}}
\newcommand{\recGapDExii}{\fpeval{round(\recDxii-\recExii,1)}}

\begin{document}

%------------------------------------------%
%     Title page
%------------------------------------------%
\begin{titlepage}
    \setstretch{1}
    \begin{center}
        \renewcommand{\thefootnote}{\fnsymbol{footnote}}
        {\setstretch{1}\LARGE\bfseries
        The American Medical Directory Panel:\\
        Comparing AI-Assisted and LLM Transcription\\
        with Manual Data Entry at Scale\footnote{We acknowledge support from the Carlsberg Foundation through a Research Infrastructure Grant (PI: Christian M. Dahl) and a Semper Ardens: Accelerate grant (PI: Anthony Wray), and from Independent Research Fund Denmark through a Research Project 1 grant (PI: Anthony Wray). We thank Michel Rasmus Jessen, Freja Vallentin Witt, \mbox{OpenAI's} ChatGPT Desktop and Codex, and Anthropic's Claude Code for research assistance, Paradigm Data Services for the manual transcription, and Harish Bindra for coordinating it. We thank Gregory Niemesh for sharing the digital images of the AMD volumes, and the librarians at Miami University (Oxford, Ohio) for photographing the volumes and pre-processing the images. We thank seminar participants at the University of Gothenburg and the Boston University Empirical Micro Reading Group, and participants at the sixth Conference of the European Society of Historical Demography in Bologna (2025), for helpful comments. The data and codebook are available on Zenodo (version \releaseVersion{}, DOI \zenodoDOI) under a CC~BY~4.0 license.}\par
        }

        \vspace{.25cm}
        % Author block: alphabetical.
        {\large
        Christian M. Dahl\textsuperscript{a} \quad
        Sophie Li\textsuperscript{a} \quad
        Rasmus Emil Nyholm\textsuperscript{a} \\[0.3em]
        Maja U. Pedersen\textsuperscript{a} \quad
        Anthony Wray\textsuperscript{a}\footnote{Corresponding author. Email: \href{mailto:wray@sam.sdu.dk}{wray@sam.sdu.dk}.}} \\[0.5em]
        {\small
        \textsuperscript{a}\,University of Southern Denmark} \\
        {\large\noteDate}
    \end{center}

    \begin{abstract}
    \begin{normalsize}
    \begin{singlespace}
    \noindent
    How closely does machine transcription agree with manual double-entry data? 
We digitize the 1912 and 1921 editions of the American Medical Directory, a national register of US physicians with roughly 140,000 and 150,000 entries in the two volumes, and compare five machine-extraction workflows with a full-count double-entered manual transcription. 
Like many historical directories, the AMD is semi-structured rather than tabular, so errors arise from misreading the layout as well as the text. 
A script that parses commercial OCR text using rules written by coding agents recovers about 97 percent of manually transcribed records with similar precision, more than a multimodal language model workflow (\geminiModelName{}) that reads the page images and the OCR text. 
When the two methods fail, they fail on partly overlapping sets of records, so reconciling their outputs raises both recall and precision to about 98 percent. 
An open-weight LLM pipeline run on a local GPU server falls between the parser and the multimodal workflow in recall, and replacing the commercial OCR model with an open-source alternative lowers its recall by only about 2 percentage points. 
As the manual transcription's own error rate is unknown, these rates measure agreement rather than accuracy. Direct processing charges for the commercial OCR and multimodal workflow were about \texteuro{}\costApiPageEurBoth{} per page, compared with \texteuro{}\costPdsPageEur{} per page for manual transcription. 
We release the data and codebook under CC~BY~4.0.

    \end{singlespace}
    \end{normalsize}
    \end{abstract}

    \begin{singlespace}
    \noindent\textbf{JEL:} C81, I11, N31, N32\par\smallskip
    \noindent\textbf{Keywords:} American Medical Directory; physicians; historical directories; OCR; large language models; transcription; record linkage; United States, 1912, 1921
    \end{singlespace}
    \setcounter{footnote}{0}
\end{titlepage}

\newpage
\FloatBarrier
\renewcommand{\thefootnote}{\arabic{footnote}}
\setlength{\floatsep}{10pt}
\setlength{\textfloatsep}{16pt plus 2pt minus 4pt}
\setlength{\intextsep}{16pt plus 2pt minus 4pt}

%------------------------------------------%
%                Main text
%------------------------------------------%
\section{Introduction}\label{sec:intro}

How closely does machine transcription agree with manual double-entry data? 
The time required to key and clean historical records limits both dataset size and the questions economic historians can investigate \citep{gutmann2018bigdata}. 
Advances in Optical Character Recognition (OCR) and large language models (LLMs) offer the potential to reduce the cost of digitizing historical sources, making large-scale transcription increasingly feasible \citep{dell2025deep, Ferrara2026Guide}. 
However, the benefits of automation depend on the accuracy of the resulting data. 
While automated transcription can reduce data-entry costs, assessing the quality of transcription still requires manual human review and raises new questions about validation. 
In particular, small character-level error rates can translate into substantial disagreement across records containing many fields \citep{correia2023digitizing}. 
Evaluating transcription accuracy therefore becomes critical.

Such evaluation raises a practical problem. 
Researchers may benchmark machine-generated data against a human-created reference dataset treated as ground truth, yet the reference dataset may itself contain transcription errors. 
Measures of agreement with a reference dataset therefore conflate machine transcription errors with errors in the benchmark, so agreement with a manual transcription does not necessarily establish accuracy \citep{DellRambachan2026Measurement}. 
This makes it difficult to determine which extraction strategy is most accurate.
As AI-assisted digitization becomes more common, distinguishing agreement from accuracy becomes increasingly important for researchers releasing and using machine-generated historical datasets.

We study this question using the 1912 and 1921 editions of the \emph{American Medical Directory} (AMD), a nationwide register of physicians in the United States published by the American Medical Association.
The AMD records physician names, locations, medical schools, graduation years, licensing year, and professional credentials, making it one of the largest and most comprehensive individual-level sources on the early twentieth-century American medical profession.
Despite its research potential, the directory remains difficult to use at scale because its information is embedded within compact, semi-structured entries rather than structured tables.

The AMD belongs to a broad class of historical sources that is common but difficult to digitize: directories and registers that list individuals under place or category headings in compact, abbreviated entries, such as city and trade directories, professional registers, and membership rolls.
Unlike tables, such sources have no fixed rows or columns.
Entries in the AMD are arranged beneath town headings and contain numerous abbreviations, irregular formatting conventions, and nested record components.
Errors can therefore arise not only from misreading individual fields but also from failing to identify record boundaries or town headings.
Because a town heading determines the location assigned to every physician listed beneath it, a single error can affect many subsequent records.
The AMD therefore allows us to distinguish errors in reading individual fields from errors in reconstructing the structure of the directory itself, a distinction that does not arise in the tabular sources studied in recent work \citep{backerperal2025llms}.

To examine both kinds of error, we compare five extraction workflows against a large double-entered manual transcription of both editions, using the same pages, record-linkage rule, and measures for each.
Precision is the share of a workflow's extracted records that link to a record in the manual transcription, and recall is the share of transcribed records that the workflow recovers; we compute both for whole records and for individual fields.
Three workflows start from the same commercial OCR: a multimodal large language model that reads page images and OCR text, a deterministic parser of OCR text whose rules were written with coding agents, and a multi-stage open-weight LLM pipeline that reads OCR text alone.
A fourth applies the same open-weight pipeline to OCR from an open-source model, which changes the OCR step (the engine and its column segmentation) and nothing downstream.
The fifth, the combined table, reconciles the multimodal LLM and parser outputs field by field and forms the basis of the public release.
Our data release includes \nDeliveryRowsxii{} and \nDeliveryRowsVzero{} rows from the two editions, including cross-reference lines, covering \nStates{} states and the District of Columbia. 
Because the case for automation rests on cost, we also compare the direct processing charges of the commercial OCR and MLLM workflow with those of the manual transcription.

Three results stand out.
First, among the four single workflows, the deterministic parser has the highest recall in both editions.
Its record-level precision and recall are \precB{} and \recB{} percent in 1921 and \precBxii{} and \recBxii{} percent in 1912, higher on both measures than the multimodal LLM in both editions (\precA{} and \recA{} percent in 1921; \precAxii{} and \recAxii{} percent in 1912) and higher on recall than the open-weight LLM pipeline.
Replacing the commercial OCR with open-source OCR in that pipeline lowers recall by \recGapDE{} percentage points in 1921 and \recGapDExii{} in 1912.
Second, when the methods fail, they fail on partly overlapping sets of records.
The combined table recovers more reference records than either input alone, with precision and recall of \precC{} and \recC{} percent in 1921 and \precCxii{} and \recCxii{} percent in 1912.
Among cells where only the parser agrees with the transcription, most for birth and graduation year are entries the MLLM workflow did not link, while most for license year and medical school are different values on linked records.
The widest gaps between methods are in county in both editions and in city in 1921. 
Third, processing charges for the commercial OCR and MLLM workflow are far below the price of manual transcription: an estimated \texteuro{}\costApiPageEurBoth{} per page, against \texteuro{}\costPdsPageEur{} per page for double-entered transcription.
This comparison excludes engineering, review, and model-development time.
Even so, digitization projects previously considered prohibitively expensive may now be feasible.

These figures should be read in light of the problem raised above: agreement with a reference dataset is not the same as accuracy.
Our reference is a double-entered transcription whose own error rate is unknown, so the reported measures quantify agreement rather than scan-verified accuracy.
Entries missed by both the machine and the transcribers are unobserved, and three of the workflows share the same OCR, so agreement among them is weaker evidence than agreement among independent readings would be.
We therefore view the paper as a methodological benchmark of alternative extraction strategies and a description of the released data.
A subsequent study will adjudicate disagreements directly against the page images and estimate accuracy using an independently verified gold-standard sample.

This paper contributes to a growing literature on large-scale historical transcription.
Existing workflows typically combine commercial OCR with pre- and post-processing to transform scanned sources into structured datasets \citep[][see \citealp{beach2023historical} for newspapers]{correia2023digitizing}.
Other projects have developed specialized layout-detection and OCR pipelines for large newspaper collections \citep{carlson2024efficient}, tabular documents \citep{Dahl2023}, and historical administrative directories \citep{fleischhacker2025enhancing}; \citet{dell2025deep} surveys these approaches.
More recently, large language models have been used to extract structured records directly from page images.
\citet{backerperal2025llms} find high agreement with a human-transcribed gold standard for historical county tables at substantially lower cost, with regression estimates of the same sign and significance, and \citet{Ferrara2026Guide} outlines a workflow that holds hand-coded validation data out of prompt development and model selection.
We add to this literature in three ways.
First, we compare multimodal, text-based LLM, and rule-based extraction on a common source and reference dataset, and we vary the OCR engine while holding the downstream pipeline fixed.
Second, we study a directory-style source whose difficulty lies in its structure as much as in its characters, and we report which records and fields the methods fail to match.
Third, we set out what agreement measures can and cannot establish when both machine-generated and human-generated transcriptions contain errors.

We also add to work that digitizes the AMD itself.
Previous studies rely on hand-keyed samples for particular states and editions \citep{moehling2019medical}, vendor double-entry transcriptions of North Carolina \citep{HollingsworthKarbownikThomassonWray2024} and of the 1906 edition for Southern and adjacent states \citep{chrisinger2024black}, and OCR extractions of single editions checked against printed physician counts \citep{smith2017digitizing} or published state totals \citep{AlsanNeberai2026QJE}.
We provide national transcriptions of two editions validated against a double-entered reference.
Appendix~\ref{app:prior} summarizes earlier AMD datasets and their reported validation procedures.

Section~\ref{sec:source} describes the directory.
Section~\ref{sec:data} describes the page images, the extraction workflows, and the public release.
Section~\ref{sec:methods} describes the validation samples, the record-linkage rule, and the agreement measures, and Section~\ref{sec:results} reports results for both editions.
Section~\ref{sec:robust-main} reports robustness checks, and Section~\ref{sec:next} discusses limitations and concludes with the planned adjudication against the page images.
Appendix~\ref{app:protocol} documents the comparison protocol.
\section{Background and Data: the American Medical Directory}\label{sec:source} 
The American Medical Directory (AMD) is a register of legally qualified and practicing physicians in the United States and Canada published every two to three years by the American Medical Association (AMA) starting in 1906 \citep{AMA1912AMD, AMA1921AMD}.
The AMA compiled the AMD volumes from an index card collection that was updated daily using reports from licensing boards and lists of newly graduated physicians \citep{AMA1912AMD}.\footnote{
To compile the first edition, the AMA drew on records from state licensing boards, medical schools, medical societies, and a direct mail canvass of physicians conducted in 1905 and 1906 \citep{JAMA1905CardIndex}. 
}
While the completeness of the AMD throughout its publication run remains an open question, the earliest editions may have omitted some physicians.\footnote{
The first edition of 1906 likely undercounts Black physicians \citep{chrisinger2024black}. 
In a few cases in each volume, an entry acknowledges that a practitioner's qualifications could not be independently verified.
} 
Among sources on the medical profession during the early twentieth century, the directory uniquely captures individual physicians and their characteristics with nationwide coverage, and thus remains an invaluable resource for studying medical education and licensing, physician labor markets, and the professionalization of medical care. 
Appendix~\ref{app:prior} provides an overview of prior hand-keyed transcriptions of AMD volumes and samples, and how these have been used in previous research. 

\begin{sidewaysfigure}
\centering
\includegraphics[width=\textwidth]{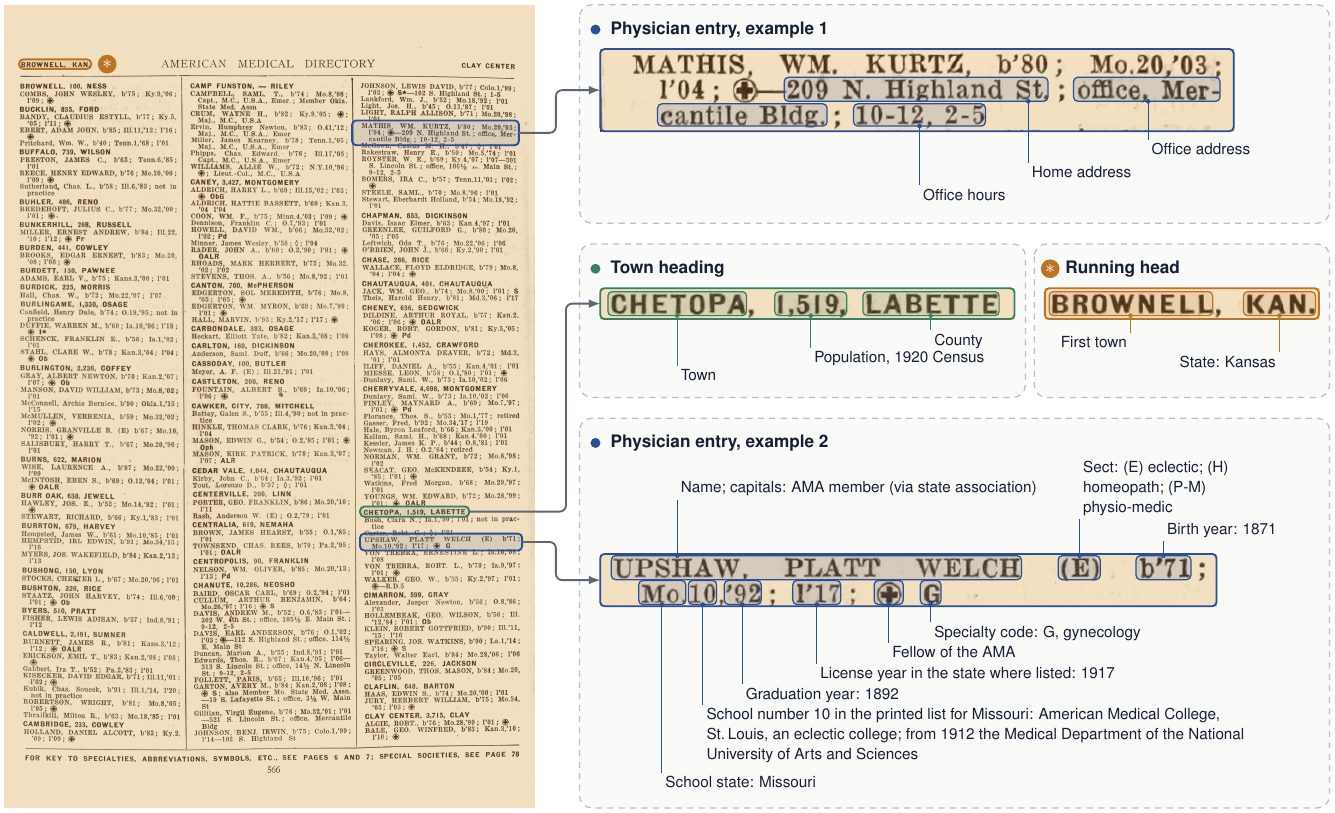}
\caption{A page of the 1921 directory and the parts of an entry}
\label{fig:entry}
\begin{minipage}{\textwidth}\footnotesize
\emph{Notes:} Page 566 of the 1921 edition (Kansas), photographed from a library copy; the photograph is straightened, evenly lit and tinted, and each enlargement is straightened separately. The green box marks a town heading, which gives the town, its population and its county; every physician listed beneath it is assigned to that town. The orange box, marked with an asterisk, is the running head, which names the first town on the page and abbreviates the state (KAN., Kansas); town headings and entries do not repeat the state. The blue boxes mark two entries, enlarged on the right. Example~1, under Chanute, shows a home address, an office address and office hours; example~2, under Chetopa, shows the name, sect, birth year, school, graduation year, license year, Fellowship mark and specialty code. Labels follow the 1921 key (Appendix~\ref{app:key}) and the list of medical colleges printed in the same volume. Other pages of the 1921 volume print a second layout, with the birth year and license year in parentheses and the Fellowship mark after the birth year. Source: \citet{AMA1921AMD}.
\end{minipage}
\end{sidewaysfigure}

Figure~\ref{fig:entry} shows a sample page of physician entries from the 1921 AMD volume, on the left, and closeups of two individual physician entries and one town heading with explanations of each field, on the right \citep{AMA1921AMD}. 
The physician directory represents one section in a vast volume that includes listings of medical officers, medical schools, medical libraries, and hospitals. 
Each volume covers physicians in every state in the continental United States and the District of Columbia, as well as the US territories and Canada.
The directory is organized hierarchically by state and town; within each state, towns are listed alphabetically. 
Each location heading reports the town name, its population, and county, and the physicians associated with that location are listed beneath it in alphabetical order. 
In some cases, a physician is listed in full under one town and appears elsewhere as a cross-reference.

The compact physician entries report medical education, credentials, professional affiliations, and details about the physician's practice. 
A typical record follows a fixed order and format: the physician's name, sect letter, two-digit birth year, medical school code and two-digit graduation year, two-digit license year, membership symbol, specialty code, home address, and office address with office hours. 
Depending on the edition and page layout, the birth year and license year may be printed in parentheses.
The exact conventions and entry order vary slightly across and within editions and do not always match the printed edition-specific key included in the directory. 
The abbreviations that appear in the AMD physician entries are explained in detail for the 1921 edition in Appendix~\ref{app:key}. 

This project draws on physician entries from every state in the continental United States and the District of Columbia in the 1912 and 1921 editions of the American Medical Directory \citep{AMA1912AMD, AMA1921AMD}.\footnote{
We exclude Canada and the US territories from the public release as they lie outside our study population. 
We further exclude entries from the ``Addendum'' section that appears at the end of printed AMD volumes -- containing corrections and late submissions -- as we lack a manual transcription of these pages to assess precision and recall (see Section~\ref{sec:reference}).
} 
In this data note, we focus on a physician's name and location, the key dates (birth, graduation, and license year), and the medical school codes. 
Many of the remaining symbols require separate processing that lies outside the scope of this study. 

\section{Digitizing the Directory: OCR and Structured Extraction Pipelines}\label{sec:data}

This section sets out the challenges with structured output extraction from the AMD, describes the page images used as source inputs and the derived OCR outputs, explains the extraction methods, and summarizes the structure of the public data release.

The semi-structured text of the physician entries and their organizational structure make the AMD difficult to digitize. 
As the town heading determines the location assigned to every physician listed beneath it, errors in identifying a heading can affect many subsequent records even when the individual entries themselves are read correctly. 
However, the relevant town may appear in the page header, in a different column on the same page, or on a separate page altogether. 
Thus, the assignment of a physician entry to the correct location requires judgment over what constitutes an entry or a town header, in addition to preserving context across images or image segments. 
These features make the AMD fundamentally different from tabular sources extracted at scale by other recent studies \citep{backerperal2025llms}. 
Successful extraction therefore requires not only accurate character recognition but also correct identification of towns and record boundaries. 
For this reason, performance depends on recovering the structure of the document as well as the contents of individual entries. 

\subsection{Page images and OCR}\label{sec:ocr}

\begin{figure}[!ht]
\centering
\includegraphics[width=\textwidth]{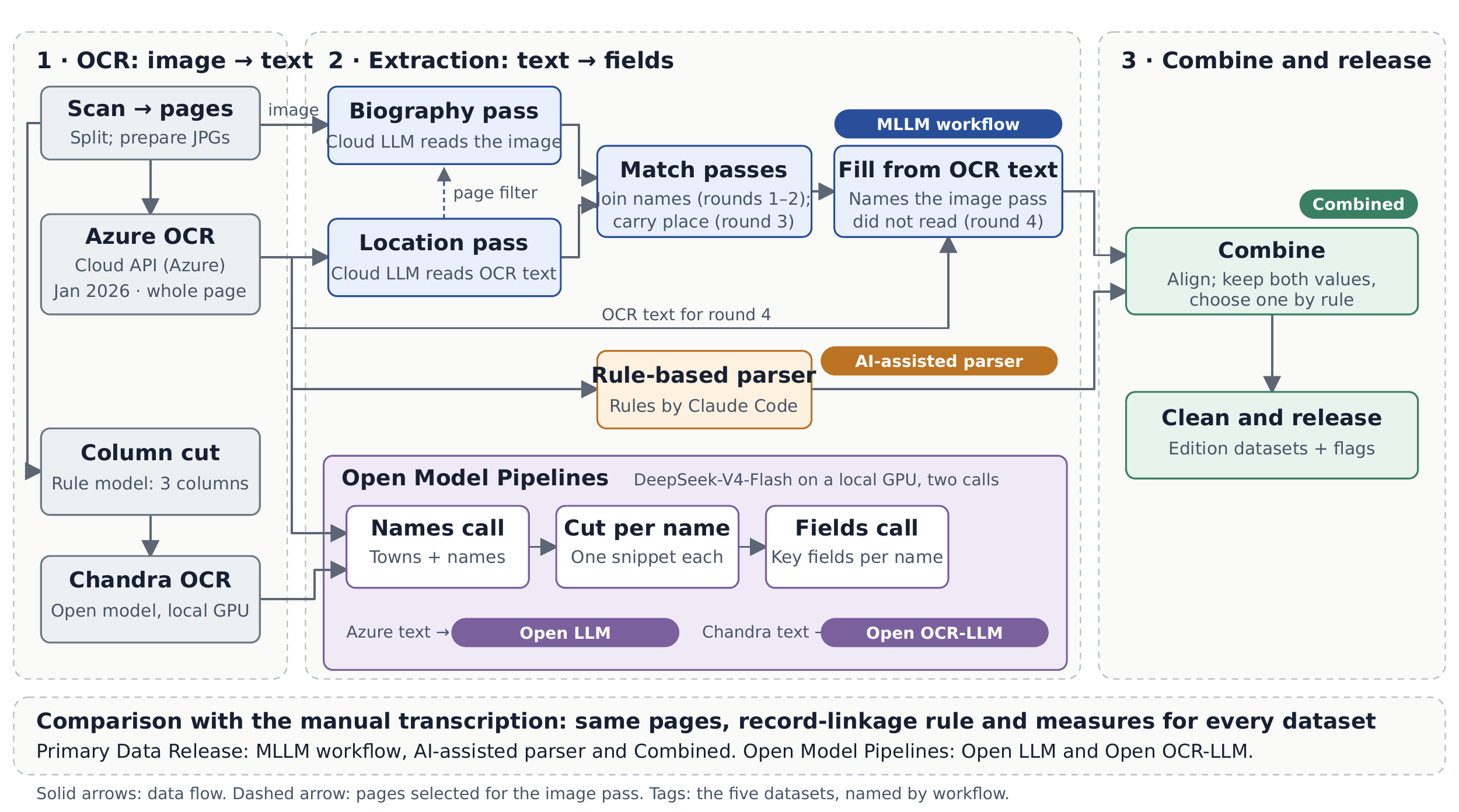}
\caption{From scans to the release tables}
\label{fig:pipeline}
\begin{minipage}{\textwidth}\footnotesize
\emph{Notes:} Solid arrows indicate data flow and the dashed arrow indicates page selection for image-based extraction. The manual transcription is used for validation and for selecting release-cleaning procedures, but it does not directly supply reconciled field values. Cross-edition linkage and record collapse occur downstream of the workflow examined in this paper. Source: project workflow documentation.
\end{minipage}
\end{figure}

Figure~\ref{fig:pipeline} provides an overview of the image pre-processing, OCR reading, structured output extraction, and validation workflows compared in this paper. 
Given that existing digitized versions of the AMD volumes available in the public domain consist of low-resolution color scans or poor-quality black and white scans, we obtained high-resolution digital images taken from physical copies of the AMD volumes as the source images for our extraction pipelines.\footnote{The images were taken by librarians at Miami University (Oxford, Ohio). We thank Gregory Niemesh for sharing the image files.} 
The pre-processing step consisted of cropping to ensure each image contained only the page itself, as well as a DPI adjustment.\footnote{
As some pages were photographed multiple times, we de-duplicate the images prior to evaluation. 
} 
These images represent a common input across all our extraction workflows.

\subsection{Five machine extraction workflows}\label{sec:readings} 

We evaluate five machine extraction workflows that each produce a structured dataset derived from page images of the directory. 
Table~\ref{tab:arms} (Appendix~\ref{app:corr}) summarizes their inputs and outputs. 
All extraction methods begin with the same underlying page images. 
Four of the workflows rely on OCR output generated by Azure Document Intelligence, a proprietary cloud API service that extracts both text and layout information (bounding box coordinates). 
We processed every page image with physician entries in the 1912 and 1921 editions during January 2026. 

\subsubsection{Multimodal language model workflow}\label{sec:gemini} 

The multimodal language model (MLLM) workflow prompts the Google Gemini API and uses the \geminiModelName{} model to extract information from the directory in two stages.\footnote{
Extraction was conducted using edition-specific prompts, and the resulting outputs from January 2026 for 1912 and February 2026 for 1921 were retained for evaluation. 
Because \geminiModelName{} was discontinued on \geminiShutdownDate, the workflow can no longer be rerun using the original model version.
}
The first stage reads the OCR text generated by Azure Document Intelligence to identify the town and county headings as well as the physician names listed beneath them. 
The second stage reads the page image and extracts the biographical fields defined by the directory key. 
A carry-forward mechanism allows location information to be filled in across page and column boundaries (Appendix~\ref{app:gemini} describes the passes and matching rounds in detail).
A deterministic linking algorithm then combines the list of towns and names from the first stage with the biographical information from the second into a single record-level dataset. 
The matching procedure blocks on page and name, while subsequent matching rounds allow for fuzzy matches on name and incorporate location information. 
A final stage uses the OCR text to fill missing biographical fields when the image-reading stage fails to extract them. This is the only stage in which OCR text contributes biographical information directly. 

\subsubsection{AI-assisted parser and combined workflow}\label{sec:parser} 

As an alternative to prompting an MLLM via API calls, we prompted a coding agent to develop a deterministic script that parses the Azure OCR text. 
The AI-assisted parser follows a set of rule-based instructions to identify document sections, locate town headings, separate individual physician entries, and extract fields according to the conventions defined in the directory key. 
For every extracted field, it retains both the original OCR text and the cleaned value.

The combined workflow aligns the outputs of the MLLM and the AI-assisted parser at the record level to produce a combined data table. 
For each field, it retains both extracted values, records the rule used to determine the release value, and preserves the alternative value for reference. 
Reconciliation uses only the two readings and fixed conventions (the medical-school code list, year-typo tables and county rules), not the manual transcription.

\subsubsection{Open-source GPU-based OCR and LLM workflows}\label{sec:ocrparser} 

As one of our objectives is to evaluate the quality of data produced by open-source models in comparison to frontier models, our final two workflows use open weights and run on a GPU server. 
We start with the same Azure OCR text as the MLLM workflow, and replace the API calls to the \geminiModelName{} model with a local deployment of DeepSeek-V4-Flash on a GPU server with temperature 0.01 to minimize variation across runs. 
The same prompts are used for both editions. 
Unlike the MLLM, the multi-stage open LLM extraction pipeline operates only on OCR text rather than page images. 
The first stage identifies town headings and physician names on each page. 
Outputs are checked automatically for obvious failures, such as implausibly short name lists, and the model call is repeated when necessary. 
The OCR text is then divided into physician-specific snippets, which form the input for the second stage. 
A second model call extracts the biographical fields defined by the directory key. 

Finally, we assess an end-to-end open-source pipeline. 
The open OCR-LLM workflow uses OCR text generated by the open-source Chandra OCR model run locally on the GPU server, rather than the Azure OCR text.
Pages are segmented into their printed columns using a table parsing algorithm prior to processing by Chandra.
These open LLM workflows produce two additional datasets that are evaluated against the same validation sample. 

\subsection{The public data release}\label{sec:release}

The public data release through Zenodo (version \releaseVersion{}, DOI \zenodoDOI) contains \nDeliveryRowsxii{} rows for the 1912 edition and \nDeliveryRowsVzero{} rows for the 1921 edition, covering physicians in every state in the continental United States and the District of Columbia.
Table~\ref{tab:desc} summarizes the public release tables.
For each edition, the release includes a primary table, the combined workflow's output, which also holds the cross-reference lines. 
Each observation is identified by edition and record identifier. 
The outputs of the open LLM and open OCR-LLM workflows are provided as separate tables (see Section~\ref{sec:results} and Appendix~\ref{app:protocol} for details). 
They are linked to the primary table through edition and record identifiers and include match-status information, but are distributed separately rather than merged into it. The agreement figures for the combined table in Section~\ref{sec:results} describe the reconciled output with its county replaced by the cleaned county; Table~\ref{tab:cleaning} reports the release table's columns as read and after cleaning. The release holds two cross-sections; it contains no identifier that links a physician across editions.

The first public release covers the state sections of the two volumes, corresponding to the material included in the manual transcription. 
The release tables include all fields evaluated in Section~\ref{sec:results}, with as-read and cleaned values for the years, medical school and county, and the workflow that found each entry. 
Additional extracted fields, including specialty, addresses and office hours, military information, printed marks (membership, fellowship, sect, and \emph{col.}), and town populations, will remain unpublished until their validation is complete. 
These fields are documented in the codebook.

The release package also contains documentation, a codebook, a manifest identifying the included editions, checksums, citation information, and a CC BY 4.0 license.
Both editions are in the public domain in the United States, which provides the basis for releasing the data tables. 
Appendix~\ref{app:key} lists the printed key elements and the release field each one feeds. 
The Zenodo record also includes a preferred citation.

\begin{table}[!ht]\centering
\begin{threeparttable}
\caption{Public Data Release: Descriptive Statistics}\label{tab:desc}
% Generated descriptive statistics for the release tables.
\begin{tabular}{lrr}
\toprule
 & 1912 & 1921 \\
\midrule
Listed physicians & 137,303 & 145,608 \\
Cross-reference lines & 1,497 & 2,432 \\
States and DC with coded rows & 49 & 49 \\
\addlinespace
\multicolumn{3}{l}{\emph{Physicians by Census region (\%)}} \\
\quad Northeast & 39,150 (28.5) & 41,787 (28.7) \\
\quad Midwest & 47,238 (34.4) & 48,185 (33.1) \\
\quad South & 39,195 (28.5) & 41,292 (28.4) \\
\quad West & 11,657 (8.5) & 14,284 (9.8) \\
\quad State not resolved to a code & 63 (0.0) & 60 (0.0) \\
\addlinespace
\multicolumn{3}{l}{\emph{Share of physicians with a value (\%)}} \\
\quad Birth year & 73.8 & 89.6 \\
\quad Medical school & 96.6 & 97.6 \\
\quad Graduation year & 96.6 & 97.6 \\
\quad License year & 78.0 & 85.2 \\
\quad City & 100.0 & 100.0 \\
\quad County & 99.8 & 99.5 \\
\bottomrule
\end{tabular}

\begin{tablenotes}\footnotesize
\item Listed rows are printed entries, not distinct people; cross-reference lines are counted on their own row. Census regions are the Bureau\textquotesingle s four regions over the \nStates{} states and the District of Columbia; a row whose state is not resolved to a code is retained and counted on its own line. A field has a value when its cleaned column is non-missing; county and medical school are counted from the values as read. Source: descriptive statistics calculated from the public data release.
\end{tablenotes}
\end{threeparttable}
\end{table}

\section{Validation}\label{sec:methods} 
\subsection{Double-entry manual transcription}\label{sec:reference}

We validate our machine-extraction workflows against a hand-keyed double-entry full-count transcription of the 1912 and 1921 AMD volumes. 
Each record was independently transcribed by two data entry workers and reconciled by a quality assurance checker. 
The manual transcription was completed during 2022, before LLM-based transcription tools were available.\footnote{The first ChatGPT model released by OpenAI launched publicly on November 30, 2022 with limited capabilities and was based on GPT-3.5.}
The transcription includes physicians located in \nStates{} US states and the District of Columbia, while excluding the addendum and address-unknown lists that appear at the back of the volumes.
The resulting spreadsheets contain one row per printed entry and one column for each recorded field. 
We treat the compiled transcription as our reference dataset for evaluation. 

\subsection{Reference population for validation}\label{sec:population}

Our reference population is the set of physician entries on printed volume pages for which we have a complete transcription. 
Table~\ref{tab:pds} (Appendix~\ref{app:corr}) summarizes the page coverage of the reference population in the 1912 and 1921 AMD volumes.  
For 1921, the manually transcribed reference dataset covers \nPagesBroad{} pages in the printed directory and \nRowsCleanedBroad{} cleaned rows. 
One complication with the manual transcription is that the data entry workers read from lower-quality scans of the 1912 and 1921 AMD volumes, rather than the high-resolution digital images.
Consequently, some scanned pages or column segments did not contain any legible content; transcribers flagged such pages or columns in the output. 
After excluding pages with incomplete reference data, around 10 to 16 percent of the total, we evaluate all machine extraction methods on a sample consisting of \nPagesFrame{} pages and \nRowsCleaned{} cleaned rows (Section~\ref{sec:measures}). 
For 1912, the manual transcription covers \nPagesBroadxii{} pages while the reference population consists of \nPagesFramexii{} pages containing \nRowsCleanedxii{} cleaned rows after excluding incomplete pages. 

\subsection{Record linkage to the validation sample}\label{sec:corr}

In the absence of a unique record identifier, we apply a record linkage procedure to merge the reference population from the manual transcription to the output of each machine-extraction workflow. 
The goal is to identify which extracted records correspond to which transcribed records. 
The resulting links form the basis for the measures of record-level agreement. 

Prior to linking, we standardize the given names, medical school codes, two-digit years, and county names in the manual transcription and the extraction outputs (See Appendix~\ref{app:protocol} for details). 
We match records using combinations of page number and physician name whenever these identifiers are unique in both datasets. 
Remaining ambiguous cases are resolved using state and city information, but only when these fields uniquely identify a match in both datasets. 
School codes and years are never used to establish a link. 
Records in the manual transcription that cannot be linked, and ties that cannot be resolved, remain unmatched and continue to count toward precision and recall denominators. 
Linkage is performed on the full set of extracted records before the analysis frame is applied. 
A field value can agree only on a linked record; values on unlinked records stay in the field denominators and count as disagreements.

\subsection{Precision and recall measures}\label{sec:measures}

The central question in this paper is how closely each machine extraction method agrees with the reference population from the manual transcription. 
We validate the extraction methods by evaluating performance at both the record and field level.
For a method $m$, let $O_m$ denote the number of extracted records, $N$ the number of reference records (on pages with complete transcription), and $L_m$ the number of linked records produced by the linkage procedure described in Section~\ref{sec:corr}.
Record-level precision is defined as $\text{Precision}_m=L_m/O_m$, that is, the share of extracted records that can be linked to a reference record. 
Record recall is defined as $\text{Recall}_m=L_m/N$, or the share of reference records recovered by the extraction method. 
A reference record is recovered when it receives a one-to-one link in the first or second pass: a page-and-name key that is unique on both sides, or, within ambiguous groups, a unique state--city pair (Section~\ref{sec:corr}).
Field-level measures are calculated similarly. 
Precision divides the number of agreeing linked values by the total number of scorable output values produced by a method, while recall divides the number of agreeing linked values by the number of scorable values in the reference dataset. 
Cells containing multiple years or multiple school codes contribute each individual value to the comparison. 
Unmatched records count as disagreements, while cases where both sides are missing are excluded.

These measures quantify agreement with the reference population used for validation rather than accuracy relative to the printed directory. 
The distinction is important because while double-entry transcription reduces keying errors, the reference data may itself contain errors.
To illustrate, consider a set of aligned records. 
Suppose the extraction method is incorrect with probability $e_m$ and the reference transcription is incorrect with probability $e_r$. 
If errors are independent and never produce the same incorrect value, the probability of agreement is $A_m=(1-e_m)(1-e_r).$
Even in this highly simplified case, observed agreement reflects both error rates. 
Recovering the extraction error rate requires information about the accuracy of the reference transcription. 
Shared errors may inflate agreement, while correlated errors invalidate the independence assumption altogether. 
High agreement therefore does not necessarily imply high accuracy.
This distinction between agreement and accuracy is central to the evaluation of machine-generated measurements \citep{DellRambachan2026Measurement}. 
Throughout the paper, the reported precision and recall statistics should therefore be interpreted as measures of agreement with the validation sample -- the share of transcribed entries recovered by an extraction method rather than the share of all printed physicians recovered from the original volumes. 
Estimating scan-verified accuracy would require an independent transcription of the same pages from which to adjudicate against the original page image and estimate the error rate of the manual transcription, which is planned for future work.

\section{Results}\label{sec:results}

This section reports the main results for the rate of agreement between the machine extraction methods and the cleaned reference population used for validation.
The measures are reported for the set of pages in each edition with a complete manual transcription (\nPagesFramexii{} pages in 1912 and \nPagesFrame{} pages in 1921). 
They quantify agreement with the reference transcription rather than scan-verified accuracy. 
All headline comparisons are based on the common precision and recall measures defined in Section~\ref{sec:measures}. 
Record-level agreement is reported in Table~\ref{tab:records}, while field-level agreement is summarized in Figure~\ref{fig:fields}. 

\subsection{Record-level agreement}\label{sec:records}

Table~\ref{tab:records} provides the main record-level comparison across extraction methods. 
Precision measures the share of output records that can be linked to a transcribed record, while recall measures the share of transcribed records recovered by a method. 
We focus on comparisons that show how agreement changes when individual components of the extraction process are altered.

\begin{table}[!ht]\centering
\small
\begin{threeparttable}
\caption{Record agreement by method, 1912 and 1921}\label{tab:records}
\footnotesize
\setlength{\tabcolsep}{3pt}
\begin{tabular}{lrrrrrrrr}
\toprule
 & \multicolumn{4}{c}{1912: \nPagesFramexii{} pages, \nRowsCleanedxii{} reference rows} & \multicolumn{4}{c}{1921: \nPagesFrame{} pages, \nRowsCleaned{} reference rows} \\
\cmidrule(lr){2-5}\cmidrule(lr){6-9}
Method & Output rows & Links & Prec.\ (\%) & Recall (\%) & Output rows & Links & Prec.\ (\%) & Recall (\%) \\
\midrule
\multicolumn{9}{l}{\emph{Primary table and its component workflows}} \\
MLLM & \nOutAxii & \nMatchedAxii & \precAxii & \recAxii & \nOutA & \nMatchedA & \precA & \recA \\
Parser & \nOutBxii & \nMatchedBxii & \precBxii & \recBxii & \nOutB & \nMatchedB & \precB & \recB \\
Combined & \nOutCxii & \nMatchedCxii & \precCxii & \recCxii & \nOutC & \nMatchedC & \precC & \recC \\
\midrule
\multicolumn{9}{l}{\emph{Open model pipelines}} \\
Open LLM & \nOutDxii & \nMatchedDxii & \precDxii & \recDxii & \nOutD & \nMatchedD & \precD & \recD \\
Open OCR-LLM & \nOutExii & \nMatchedExii & \precExii & \recExii & \nOutE & \nMatchedE & \precE & \recE \\
\bottomrule
\end{tabular}
\begin{tablenotes}\footnotesize
\item This table reports record-level precision and recall for each of the five machine extraction workflows and each AMD volume (1912 and 1921) compared to the manual transcription. The reference population for validation is the set of pages with complete manual transcription coverage. Measures reflect agreement with the cleaned validation sample. Recall uses the cleaned reference-row count, while precision uses each method's output-row count. All workflows count printed cross-reference lines as records, in both denominators and with the same linkage key. 
\item Primary table and its component workflows: multimodal language model (MLLM), \geminiModelName{} on the page image and the Azure OCR text; AI-assisted parser, a deterministic parser on the Azure OCR text; Combined, the parser and MLLM outputs merged field by field.
\item Open model pipelines: Open LLM, a multi-stage LLM pipeline (DeepSeek-V4-Flash) on the Azure OCR text; Open OCR-LLM, the same pipeline on OCR text from an open-source model (Chandra).
\end{tablenotes}
\end{threeparttable}
\end{table}

\textit{Overall performance:} Three findings stand out. First, the AI-assisted parser consistently outperforms the MLLM workflow. 
Second, combining the MLLM and parser outputs produces the highest overall recall. 
Third, agreement is broadly similar across the 1912 and 1921 editions.

\textit{AI-assisted parser versus MLLM workflow:} The clearest contrast is between the deterministic AI-assisted parser and the MLLM workflow. In both editions, the parser links more reference records and does so with higher precision. For 1921, the parser achieves precision and recall of \precB/\recB\%, compared with \precA/\recA\% for the MLLM workflow. The same pattern appears in 1912, where the parser reaches \precBxii/\recBxii\%, compared with \precAxii/\recAxii\%. The precision advantage is between two and three percentage points in both editions, while recall is \recGapBA{} percentage points higher in 1921 and roughly three percentage points higher in 1912. For this source, the deterministic parser achieves both higher precision and higher recall than the multimodal extraction workflow.

\textit{Image-based and text-based extraction within the MLLM workflow:} The MLLM workflow combines two stages implemented with \geminiModelName{}. The location stage operates on OCR text, while the biography stage operates on the page image itself. Table~\ref{tab:stages} shows that the two stages achieve remarkably similar linkage rates. In 1921, the text-based stage links \recStageLoc\% of the transcribed rows at \precStageLoc\% precision, while the image-based stage links \recStageBio\% at \precStageBio\% precision. The differences are less than half a percentage point in precision and about one percentage point in recall. These figures do not provide a pure comparison between text and image inputs. The two stages perform different tasks: the OCR-based stage identifies town headings and physician names, while the image-based stage extracts biographical information.

\textit{Combining extraction methods:} The combined table achieves the highest recall in both editions. In 1921 it reaches \precC/\recC\%, compared with \precB/\recB\% for the parser. In 1912 it reaches \precCxii/\recCxii\%, compared with \precBxii/\recBxii\%. The gains are modest: recall rises by 0.83 percentage points in 1921 and 0.44 in 1912, and precision by 1.32 and 0.18. The two methods therefore fail on partly overlapping sets of records, although most of the reference records that the parser does not link remain unlinked in the combined table. The largest field-level change is in county (Section~\ref{sec:fields}), where most of the gain comes from a county-cleaning rule that we selected by comparison with the transcription, not from reconciliation.

\textit{Open LLM versus AI-assisted parser:} The comparison between the open LLM workflow and the AI-assisted parser contrasts two OCR-based extraction workflows: one based on a multi-stage LLM pipeline and one based on deterministic parsing rules. In 1921, the open LLM workflow achieves precision \precGapBD{} percentage points below the parser and recall \recGapBD{} percentage points lower. In 1912, precision is about 0.19 percentage points higher and recall is about one and a half percentage points below the parser (\recDxii\% compared with \recBxii\%). This comparison should be interpreted descriptively rather than causally. The two workflows differ not only in their extraction models but also in prompting strategy and document handling.

\textit{Changing the OCR source:} The open OCR-LLM workflow applies the same code, model, and prompts as the open LLM workflow, but its OCR comes from an open-source model run on pages that are first segmented into columns. The contrast therefore changes the OCR step, the engine and the column segmentation together, and nothing downstream. Recall falls by \recGapDE{} percentage points in 1921 and \recGapDExii{} in 1912, and precision by \fpeval{round(\precD-\precE,1)} and \fpeval{round(\precDxii-\precExii,1)} (Table~\ref{tab:records}).

\textit{Comparisons across editions:} Precision and recall are broadly similar across the 1912 and 1921 editions, generally within about a point of the other edition for the released workflows. The largest exception is the open LLM workflow, whose precision is about 1.7 percentage points higher in 1912 (\precDxii\%) than in 1921 (\precD\%). The comparison does not allow us to determine whether this difference arises from characteristics of the page images themselves or from edition-specific processing decisions. The editions differ in their printed keys and, for the MLLM workflow, in the prompts. The open model pipelines use the same prompts for both editions, so their difference across editions is not a prompt effect.

\subsection{Field-level agreement}\label{sec:fields}

\begin{figure}[!p]\centering
\includegraphics[width=\textwidth]{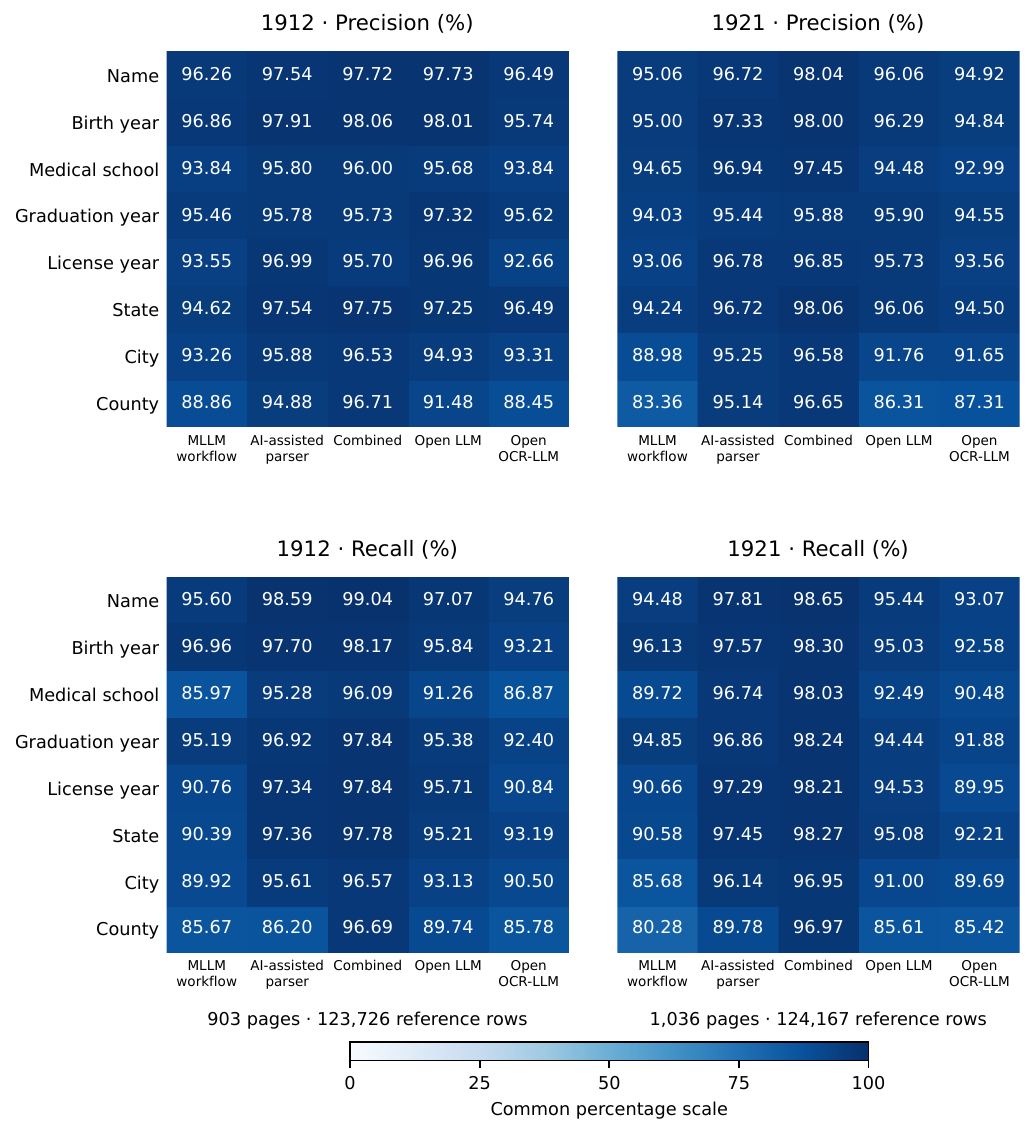}
\caption{Field agreement by method, 1912 and 1921}\label{fig:fields}
\begin{minipage}{\textwidth}\footnotesize
\emph{Notes:} \pdslabel{} Same frames, links and scorer as Table~\ref{tab:records}. Precision divides agreeing linked values by all scorable output values; recall divides them by all scorable reference values. Unmatched values count as failures; missing/missing is excluded. Both editions and measures use the same 0--100\% color scale, with percentages rounded to two decimals. Name and, for second-pass links, state/city agreement are linkage-conditioned. Years are two-digit tokens, school codes syntactically normalized, and places normalized for formatting.  The combined table's county uses the checked county-cleaning rule on fixed links; its other fields use the reconciled reading. Table~\ref{tab:fields} retains the exact displayed percentages; Table~\ref{tab:fieldcounts} gives the integer counts for both editions.
\end{minipage}
\end{figure}

Record-level agreement masks important differences across fields. 
Figure~\ref{fig:fields} therefore examines agreement separately for names, locations, medical school codes, and other biographical characteristics. 
Exact percentages are reported in Table~\ref{tab:fields} (Appendix~\ref{app:fields}). 

The broad pattern mirrors the record-level results. 
Across most biographical fields, the AI-assisted parser and the combined table achieve the highest levels of agreement with the reference transcription. 
In 1921, agreement for these methods ranges between \rangeFieldsBC{} percent across the evaluated biographical fields. 
The corresponding range is \rangeFieldsA{} percent for the MLLM workflow and \rangeFieldsD{} percent for the open LLM workflow. 
For the open LLM workflow, medical school is the weakest of the four biographical fields, with recall of \fieldSchoolRD\%. 
The same pattern appears in 1912, although field-level agreement varies somewhat more across fields. 
Medical school remains one of the most challenging fields for all methods and is again the weakest of the four biographical fields for the open LLM workflow (\fieldSchoolRDxii\%). 

The largest differences across methods occur in place information. 
City agreement reaches \rangeCityBC{} percent for the AI-assisted parser and the combined table in 1921, compared with \rangeCityA{} percent for the MLLM workflow. 
For 1912, the corresponding ranges are \rangeCityBCxii{} and \rangeCityAxii{} percent. 
County shows the largest gain for the combined table, but most of it comes from county cleaning. 
In 1921, county precision and recall are \fieldCountyB{} for the parser and \fieldCountyC{} for the combined table. In 1912, the corresponding values are \fieldCountyBxii{} and \fieldCountyCxii. 
The combined table's county is scored after the county-cleaning rule. 
As read, its 1921 figures are \cleanCountyAsRead{} (Table~\ref{tab:cleaning}). 
Reconciliation alone therefore raises county recall over the parser by about one point and lowers precision, and the cleaning rule accounts for the rest. %Because we selected that rule by comparison with the transcription, the cleaned figure is not an out-of-sample result. 

\subsection{Sources of disagreement}\label{sec:disagree}

Understanding why methods disagree is as important as measuring how often they disagree. 
In many cases, disagreements arise not because a field value differs, but because a record is not linked: no record on the other side has the same page and name key, or the match is ambiguous. 
We currently have the most detailed diagnostic information for the 1921 edition. 
For the combined table, recall losses fall into three main groups: \nPdsNoKeyC{} transcribed records whose page-and-name key has no corresponding extracted record, \nPdsAmbiguousC{} records whose key could not be resolved during second-pass matching, and \nPdsMissingKeyC{} records lacking a usable name or page identifier. 
Precision losses arise from \nOutNoKeyC{} machine-generated records with no corresponding transcription record and \nOutAmbiguousC{} records that remain ambiguous after matching. 
The available diagnostics cannot determine whether these discrepancies originate in the extraction process or in the reference transcription itself. 
Resolving these cases requires direct inspection of the page images. 

Table~\ref{tab:directions} (Appendix~\ref{app:fields}) counts the disagreements at the field level, and Table~\ref{tab:strata} (Appendix~\ref{app:corr}) reports record agreement separately for pages where the transcribers flagged many entries as uncertain, and recall separately for the entries they flagged. 
Among observations where only one of the two component methods behind the combined table agrees with the reference transcription, the parser agrees more frequently for all four evaluated fields. 
For birth year and graduation year, most of this advantage reflects missing records in the MLLM output. 
Missing entries account for \disBirthYearParserRightRowAbsent{} of \disBirthYearParserRight{} disagreements in birth year and \disGradYearParserRightRowAbsent{} of \disGradYearParserRight{} disagreements in graduation year. 
For license year and medical school, missing records explain a smaller share of disagreements. 
They account for only \disLicenceYearParserRightRowAbsent{} of \disLicenceYearParserRight{} license-year discrepancies and \disSchoolParserRightRowAbsent{} of \disSchoolParserRight{} school-code discrepancies. 
In the remaining cases, both outputs contain a record, but one workflow extracts a different field value from the reference transcription. 
In Table~\ref{tab:strata}, record agreement is also lower, in all five workflows, on pages where the transcribers flagged at least five percent of entries as uncertain.

\subsection{Cost}\label{sec:cost}

The commercial OCR and MLLM workflow cost an estimated \texteuro{}\costApiPageEurBoth{} per page in direct processing charges (Gemini plus Azure OCR), roughly \costApiRatioRounded{} times less than manual double-entry transcription, which cost \texteuro{}\costPdsPageEur{} per page. 
These figures cover processing charges only and leave out engineering, review, model development, coding-agent assistance and the local computing used to run the open model pipelines, which also serves other project work. 
The machine outputs and the manual transcription do not cover exactly the same pages and fields, so this is not an end-to-end productivity comparison (Appendix~\ref{app:cost}).

\FloatBarrier

\section{Robustness}\label{sec:robust-main}

The ordering of the AI-assisted parser and the MLLM workflow is robust to the choice of reference population, links, and page restrictions. 
Using all manually transcribed pages, including those the transcribers flagged as having incomplete coverage, moves recall by about 0.3 points or less in both editions and 1912 precision by at most 0.16 points. 
It lowers 1921 precision by about 1.6 points for every workflow, because machine rows for entries the transcribers skipped count as errors (Appendix~\ref{app:robust}). 
Record agreement is lower on pages where the transcribers flagged at least five percent of entries as uncertain, in all five workflows (Table~\ref{tab:strata}, Appendix~\ref{app:corr}). 
Using first-pass links only shifts 1921 record agreement by about 0.2 points or less (Appendix~\ref{app:corr}). 
Dropping the pages used in parser development changes precision and recall by at most 0.23 percentage points for every workflow in both editions, and the ordering of workflows by recall stays the same (Appendix~\ref{app:robust}).

\section{Conclusion}\label{sec:next} 
How closely does machine transcription agree with double-entered manual transcription at full-count scale? 
This paper compares five machine extraction workflows against a double-entered full-count transcription of the 1912 and 1921 volumes of the American Medical Directory, holding roughly 140,000 and 150,000 entries of physician biographies, respectively. 
Machine extraction agrees closely with the double-entered manual transcription. 

In 1921 the AI-assisted parser recovers \recB\% of the manually transcribed entries and the MLLM workflow \recA\%; in 1912 the figures are \recBxii\% and \recAxii\%. 
Precision is similar, at \precB\% and \precA\% in 1921 and \precBxii\% and \precAxii\% in 1912. 
The open LLM workflow falls between the MLLM workflow and the parser in recall. 
The open LLM workflow reaches \recD\% recall in 1921 and \recDxii\% in 1912, and replacing the commercial OCR with open-source OCR in the open OCR-LLM workflow costs \recGapDE{} percentage points of recall in 1921 and \recGapDExii{} in 1912. 
The practical lesson is to combine workflows. 
The MLLM workflow and the AI-assisted parser fail on partly overlapping sets of records, and the combined table reaches \recC\% recall in 1921 and \recCxii\% in 1912, above any single workflow. 
Direct processing charges for the commercial OCR and MLLM workflow are an estimated \texteuro{}\costApiPageEurBoth{} per page, against \texteuro{}\costPdsPageEur{} per page for manual transcription, roughly \costApiRatioRounded{} times less before engineering, review and hardware.

As we have emphasized, the reported precision and recall measures quantify agreement with the reference population, and do not measure scan-verified accuracy. 
Entries missed by both the extraction methods and the manual transcription are unobserved, so recall is defined relative to the reference transcription rather than to the printed directories themselves. 
Several questions therefore remain beyond the scope of the present comparison, including the accuracy of the reference transcription itself. 
The extraction methods are also not fully independent. 
The MLLM workflow, the AI-assisted parser, the combined table and the open LLM workflow all rely on the same Azure OCR text of the underlying page images. 
Agreement among these methods therefore provides less evidence than would agreement among fully independent readings. 
The disagreements examined in Section~\ref{sec:disagree} provide useful diagnostic information, but they do not establish which source is correct. 
Resolving such cases requires adjudication against the original page images. 

Several extensions are planned. 
First, we will construct a scan-verified evaluation sample based on direct adjudication against the original page images, allowing accuracy to be assessed rather than inferred from agreement. 
Second, we will classify disagreements by extraction stage and apply the construct-versus-observation framework of \citet{DellRambachan2026Measurement} to selected variables. 
Third, we will evaluate additional extraction workflows designed to provide a more independent comparison.

%------------------------------------------%
%                References
%------------------------------------------%
\newpage
\FloatBarrier
\begin{singlespace}
\bibliographystyle{econ}
\bibliography{amd_data_note_2026}
\end{singlespace}

%------------------------------------------%
%              Appendix
%------------------------------------------%
\newpage
\appendix
\numberwithin{figure}{section}
\numberwithin{table}{section}
% Each appendix starts on a new page and keeps its floats: flush floats, then break the page, before every section.
\let\origsection\section
\renewcommand{\section}{\FloatBarrier\clearpage\origsection}
% Appendices in the order the main text first cites them.
\section{Prior transcriptions of the AMD}\label{app:prior}

Because successive editions repeat a recurring core of fields for the licensed profession, the AMD supports three kinds of study that few other historical sources allow: the supply and location of physicians in a place and its change over time \citep{moehling2019medical, MountinPennellNicolay1942General}; the training of physicians, through the medical school code and graduation year; and the careers of groups the directory marks, Black physicians through the \emph{col.}\ label \citep{chrisinger2024black} and members and specialists through the membership marks and specialty codes \citep{AlsanNeberai2026QJE}.

Table~\ref{tab:prior} lists prior AMD transcriptions, including hand-keyed subsets and national OCR extractions of single editions, 1918 for \citet{smith2017digitizing} and 1950 for \citet{AlsanNeberai2026QJE}.
\citet{moehling2019medical} used hand transcriptions of four editions for four states to study physician labor markets. 
Our own earlier work used a vendor double-entry transcription of the AMD for North Carolina from every edition between 1912 and 1942 \citep{HollingsworthKarbownikThomassonWray2024}. 
\citet{AlsanNeberai2026QJE} digitized and OCR'd the 1950 edition for membership and specialists and checked state totals against published aggregates. 
Other studies use a state, a city or sample \citep{Staines2015, CatillonCutlerGetzen2018, MullallyMacDonald2017, McNamaraHassingerMitchell1955, Weisz2006RegulatingSpecialists, Schafer2014}. 
\citet{andrews2023physician} uses AMD aggregates and a state directory, not AMD microdata.

The closest precedent to our validation samples is \citet{chrisinger2024black}, which used a double-entry manual transcription for eighteen Southern and adjacent states and the District of Columbia. 
The article analyzes \nChrisingerEntries{} entries, including \nChrisingerLabelled{} physicians labeled \emph{col.}\ and \nChrisingerBlackSchool{} unlabeled graduates of Black medical schools, and its dataset is publicly deposited \citep{ChrisingerOxfordFigshare2023v2}. %The article concedes an undercount. We did not establish a numerical keying-error rate from the main text; the supplementary validation methods require separate examination.
\citet{smith2017digitizing} provides a machine-transcription precedent with reported record checks. 
The 1918 prototype uses commercial OCR and two parsing programs, keeps \smithEntriesKept{} percent of single-line entries, varying by state, and reports more than \smithEntriesPerfect{} percent of manually checked entries as perfect. 
It also compares capture with printed counts for the largest cities.

A series of articles in \emph{Public Health Reports} by Mountin, Pennell and coauthors examines physician location and movement between 1923 and 1938 \citep[for example][]{MountinPennellNicolay1942General}. Their titles and records show the subject but not how physicians were linked across editions.

\begin{table}[!ht]\centering
\footnotesize
\setlength{\tabcolsep}{4pt}
\begin{threeparttable}
\caption{Prior transcriptions of the American Medical Directory}\label{tab:prior}
\begin{tabular}{p{3.0cm}p{1.8cm}p{2.6cm}p{2.4cm}p{3.2cm}p{1.4cm}}
\toprule
Study & Editions & Coverage & Method & Validation reported & Public \\
\midrule
\citet{moehling2019medical} & 1909, 1914, 1918, 1921 & CA, MS, NY, NC & hand keying & --- & --- \\
\citet{HollingsworthKarbownikThomassonWray2024} & 1912--1942 & North Carolina & vendor double entry\tnote{a} & --- & yes, deposited \\
\citet{AlsanNeberai2026QJE} & 1950 by OCR; earlier editions cited & national (membership, specialists) & OCR & state totals vs published aggregates & --- \\
\citet{smith2017digitizing} & 1918 & national pilot & OCR, two parsers & printed per-city counts; \smithEntriesKept\% of single-line entries kept; $>$\smithEntriesPerfect\% of checked entries perfect & --- \\
\citet{chrisinger2024black} & 1906 & 18 Southern and adjacent states, DC & vendor double entry & likely undercount reported & yes, deposited \\
\citet{MountinPennellNicolay1942General} & 1923, 1938 & national & Public Health Service tabulation & --- & in print \\
This note & 1912, 1921 & 48 states and DC & OCR; MLLM, AI-assisted parser, open LLM and open OCR-LLM workflows & agreement with a double-entered transcription on its completely transcribed pages & yes, Zenodo, CC~BY \\
\bottomrule
\end{tabular}
\begin{tablenotes}\footnotesize
\item[a] Our own earlier work; the vendor reports double entry.
\item[] --- : not examined for this note.
\end{tablenotes}
\end{threeparttable}
\end{table}

\section{Scoring protocol}\label{app:protocol}

This appendix collects the rules under which every figure in Section~\ref{sec:results} was produced.

\paragraph{Record linkage.}
Each workflow is linked to the validation sample separately, with the same rule in both editions.
The first pass matches printed-page and name keys that occur exactly once on each side.
Within page-and-name groups that remain ambiguous, the second pass matches state--city pairs that are valid and unique on both sides.
Birth, graduation and license years and the school code never select a link.
Unmatched rows and unresolved ties remain in both denominators, and links are fixed over the whole edition before any page restriction, so name agreement is conditioned on the rule. The linkage key is the printed page and the name as a multiset of tokens: names are normalized (Unicode composition, case folding), split into alphanumeric tokens and sorted, with no fuzzy repair, and a blank name gives no key. Years and school codes are never keys.
Table~\ref{tab:corr} (Appendix~\ref{app:corr}) gives the 1921 figures after the first pass alone.

\paragraph{Scorer.}
All headline tables use a common scorer.
The scorer applies the same normalization to city and county on both sides before linking: it removes accents, the \emph{St.}/\emph{Mt.}/\emph{Ft.}\ abbreviations, a trailing \emph{County}, punctuation and spacing, and nothing else.
Field precision divides agreeing linked values by all scorable output values and field recall by all scorable reference values; an unmatched row's values count as failures; a blank on both sides never counts as agreement; years are compared as two-digit tokens and school codes after the same syntactic normalization on both sides.

\paragraph{Cleaned reference.}
The scorer reads a \emph{cleaned} version of the transcription, which normalizes school codes and two-digit years to the forms the key defines and does not touch names or places.
Its largest effect is on the school code, where the transcribers' varied spellings of one code would otherwise count as disagreement.
That difference reflects how the reference represents a value, not the extraction.

\paragraph{Frames.}
The analysis frame of an edition is the set of transcribed printed pages \emph{without known incomplete coverage}: pages on which the compiled transcription records no ignored-row placeholder, including the pages used to tune the parser, the pages drawn into a sampling design and the diagnostic pages (Appendix~\ref{app:robust} reports the frame without them).
For 1921 that is \nPagesFrame{} pages and \nRowsCleaned{} cleaned rows out of \nPagesBroad{} transcribed page keys, after dropping the \nPagesRobIncomplete{} pages with an ignored-row placeholder.
For 1912 it is \nPagesFramexii{} pages and \nRowsCleanedxii{} rows out of \nPagesBroadxii{} page keys, after dropping the \nPagesRobIncompletexii{} pages with an ignored-row placeholder.
Table~\ref{tab:pds} (Appendix~\ref{app:corr}) sets the two editions side by side; Table~\ref{tab:corr} gives the 1921 figures on the broad frame.

\paragraph{Evaluated inputs.}
The MLLM workflow is the matched output of the stored biography and location passes; the 1921 model passes date from February 2026, and the 1912 output is the one used in the tables prepared for release.
The AI-assisted parser is the parser\textquotesingle s raw candidate list for each edition.
The combined table is the reconciled output in each edition, with the checked cleaned county substituted for the as-read county. The paired county checks retain exactly the same links and all non-county fields.
The open LLM workflow is open-model-pipeline output for 1921 (\nEvalRowsDVzero{} evaluated rows, of \nOutputRowsDVzero{} output rows with \nDupPageRowsDVzero{} duplicate-page rows dropped) and 1912 (149,471 rows).
The open OCR-LLM workflow is the output of the same open model pipeline, with the same model and prompts, on the open-source OCR text of each edition.
The open LLM and open OCR-LLM outputs carry their own page identifiers, which a reviewed page bridge maps to printed pages in both editions.
Table~\ref{tab:arms} (Appendix~\ref{app:corr}) lists the inputs of each workflow.

\section{The printed key for 1921}\label{app:key}

\paragraph{What an entry holds.}
The body is arranged by state, then by town, with the town's population and county on a heading line and the physicians beneath it in alphabetical order, in narrow columns.
A physician listed in full under one town may be printed under another as a cross-reference line.
Each entry packs its fields into a fixed order of abbreviations, decoded by an edition-specific printed key; the front-matter key is not in our scans, and we read it from an Internet Archive copy \citep{AMA1921AMD}.
Table~\ref{tab:key} lists the printed elements and the release field each feeds.
The license year is the year of license in the state where the physician is listed, so a mover carries a new license year in the next edition; a missing year has several printed meanings, most with their own mark, and some states print no year by convention.
The release tables carry the year only, not which reason the mark gave.
The address element \texttt{R.D.}, the physician count, the town line and the cross-reference are layout elements rather than key definitions; we read their meanings from the entries themselves.

\begin{table}[!ht]\centering
\small
\begin{threeparttable}
\caption{The 1921 printed key and release fields}\label{tab:key}
\begin{tabular}{p{3.4cm}p{6.6cm}p{4.6cm}}
\toprule
Printed element & Meaning in 1921 & Release field \\
\midrule
Name in capitals & membership in the AMA through a constituent state association and component county society & withheld \\
AMA mark, letter-coded before OCR, after the license year\tnote{c} & Fellowship in the AMA; commissioned medical officers on active service are Fellows\tnote{a} & withheld \\
\texttt{(E)}, \texttt{(H)}, \texttt{(P-M)} after the name & eclectic, homeopath, physio-medic & withheld \\
\texttt{b'75} & born 1875 & birth year, as read and cleaned \\
\texttt{Md.7,'09} & school key number (state abbreviation and number) and year of graduation & school code (cleaned to the key's form), graduation year \\
$\lozenge$ & information received does not show graduation from a medical college\tnote{a} & not captured \\
$\triangle$ & graduate of an institution not listed as a medical college by the Council on Medical Education and Hospitals\tnote{a} & not captured \\
\texttt{l'20} & year of license in the state where located & license year \\
\texttt{l}\dag & licensed, year not known & not captured \\
\ddag & information received does not show license in the state where located & not captured \\
circle under a cross & in practice in the state where located before passage of the present law, and licensed by filing a diploma or by years of practice & not captured \\
Specialty code (\texttt{S}, \texttt{Oph}, \texttt{Pd}, \ldots), \texttt{*} & self-reported interest; with the asterisk, practice claimed limited to that branch; printed on request only & withheld \\
\texttt{R.D.}, \texttt{R.D.1} & rural-delivery address element, never a specialty & withheld with the addresses \\
\texttt{(col.)} after the name & the Association's label for a Black physician\tnote{b} & withheld \\
\texttt{NUMBER OF PHYSICIANS, $n$} under a town line & count of physicians listed under that town & not released \\
Town line \texttt{TOWN, population, COUNTY} & population and county of the town & county, as read and cleaned; population withheld \\
\texttt{(See Chicago)} & cross-reference to the full entry under another town & cross-reference line in the release table \\
ADDENDA; index symbols & physicians received too late to classify; corrections; reports of deaths & not released \\
\bottomrule
\end{tabular}
\begin{tablenotes}\footnotesize
\item[a] Carried from the 1918 key or the 1921 addenda key; the 1921 front-matter key is not in our scans.
\item[b] Not defined in any key we have seen; the label appears only in entries, and its reading follows \citet{chrisinger2024black}.
\item[c] Or after the birth year on pages printed in the second layout.
\end{tablenotes}
\end{threeparttable}
\end{table}

\section{Linkage sensitivity and stage decomposition}\label{app:corr}

Appendix~\ref{app:protocol} states the linkage rule, scorer, frames and inputs. Table~\ref{tab:pds} gives the coverage of the validation samples (Section~\ref{sec:reference}) and Table~\ref{tab:arms} the five extraction workflows (Section~\ref{sec:readings}).

\begin{table}[!ht]\centering
\begin{threeparttable}
\caption{Coverage of the validation samples, 1912 and 1921}\label{tab:pds}
\begin{tabular}{p{7.5cm}rr}
\toprule
& 1912 & 1921 \\
\midrule
Printed-page keys transcribed (broad frame) & \nPagesBroadxii & \nPagesBroad \\
Cleaned rows, broad frame & \nRowsRobAllxii & \nRowsCleanedBroad \\
Pages without known incomplete coverage (analysis frame) & \nPagesFramexii & \nPagesFrame \\
Cleaned rows, analysis frame & \nRowsCleanedxii & \nRowsCleaned \\
Analysis pages seen while tuning the parser & \nPagesExposedTuningxii & \nPagesExposedTuning \\
Analysis pages in the frozen design sample & \nPagesExposedDesignxii & \nPagesExposedDesign \\
Analysis pages used in diagnostics & \nPagesExposedDiagnosticxii & \nPagesExposedDiagnostic \\
\midrule
Fields transcribed & \multicolumn{2}{p{6cm}}{name, birth year, school, graduation year, license year, city, county, state, population} \\
Not transcribed & \multicolumn{2}{p{6cm}}{addenda; address-unknown list; count lines (inconsistently)} \\
\bottomrule
\end{tabular}
\begin{tablenotes}\footnotesize
\item Sources: comparison results and validation-sample documentation for each edition.
\end{tablenotes}
\end{threeparttable}
\end{table}

\begin{table}[!ht]\centering
\small
\begin{threeparttable}
\caption{The five extraction workflows}\label{tab:arms}
\setlength{\tabcolsep}{4pt}
\begin{tabular}{p{1.8cm}p{2.4cm}p{2.8cm}p{2.7cm}p{3.2cm}}
\toprule
Workflow & Input & Model or rules & Fields produced & Evaluated output (1921) \\
\midrule
MLLM & page image (biography); OCR text (location) & \geminiModelName, prompts per edition; page matching; OCR enhancement & name, birth year, school, graduation year, license year, specialty, city, county, state & biography and location runs of February 2026; matching run of October 2026 \\
Parser (AI-assisted) & OCR text & deterministic parser; same rules and dictionary in both editions & all of the above, plus population, membership and military flags, cross-reference lines & raw candidate list (Appendix~\ref{app:protocol}) \\
Combined & MLLM and AI-assisted parser & per-field reconciliation rules; release cleaning; 48-state rule & the parser\textquotesingle s fields with chosen values and per-field provenance & reconciled output; checked cleaned county on fixed links \\
Open LLM & OCR text & multi-stage LLM extraction (open model pipeline) & name, birth year, school, graduation year, license year, city, county, state & 1921: \nEvalRowsDVzero{} evaluated rows (of \nOutputRowsDVzero{} output rows, with \nDupPageRowsDVzero{} duplicate-page rows dropped) \\
Open OCR-LLM & OCR text of the page images by an open-source model & open model pipeline (same code, model and prompts) & as Open LLM & scores in Tables~\ref{tab:records} and~\ref{tab:fields} \\
\bottomrule
\end{tabular}
\begin{tablenotes}\footnotesize
\item The MLLM workflow, the AI-assisted parser, the combined table and the open LLM workflow consume the same stored OCR output; the MLLM workflow also reads page images for biographical fields. The open OCR-LLM workflow consumes a separate OCR of the same page images and shares no OCR text with the others. Specialty and symbol flags are neither scored here nor included in the public release.
\end{tablenotes}
\end{threeparttable}
\end{table}

Table~\ref{tab:corr} reports the record comparison of Section~\ref{sec:records} for both editions under the first pass alone and after the second pass, and on the broad frame of every transcribed page key (\nPagesBroadxii{} pages in 1912 and \nPagesBroad{} in 1921), which includes the pages with known incomplete coverage.
The broad frame lowers precision for every workflow: on a page the transcribers covered only in part, a machine row for a physician they skipped counts against the extraction.
Table~\ref{tab:stages} decomposes the MLLM workflow by stage on the analysis frames; Appendix~\ref{app:gemini} describes the stages.
Table~\ref{tab:strata} splits record agreement by the share of each page's entries that the transcribers flagged as uncertain and, for recall, by each entry's own flag; it is the evidence for Section~\ref{sec:disagree}.

\begin{table}[!ht]\centering
\small
\begin{threeparttable}
\caption{Record agreement by link pass and page frame, 1912 and 1921 (precision / recall, \%)}\label{tab:corr}
\footnotesize
\setlength{\tabcolsep}{3pt}
\begin{tabular}{lrrrrrr}
\toprule
 & \multicolumn{3}{c}{Links and rows, analysis frame} & \multicolumn{2}{c}{Analysis frame} & Broad frame \\
\cmidrule(lr){2-4}\cmidrule(lr){5-6}\cmidrule(lr){7-7}
Workflow & First-pass links & Second pass adds & Output rows & First pass & Second pass & Second pass \\
\midrule
\multicolumn{7}{l}{\emph{1912} (\nPagesFramexii{} analysis pages)} \\
MLLM & \nStrictAxii & \nSecondAxii & \nOutAxii & \precAstrictxii{} / \recAstrictxii & \precAxii{} / \recAxii & \precAbroadxii{} / \recAbroadxii \\
Parser & \nStrictBxii & \nSecondBxii & \nOutBxii & \precBstrictxii{} / \recBstrictxii & \precBxii{} / \recBxii & \precBbroadxii{} / \recBbroadxii \\
Combined & \nStrictCxii & \nSecondCxii & \nOutCxii & \precCstrictxii{} / \recCstrictxii & \precCxii{} / \recCxii & \precCbroadxii{} / \recCbroadxii \\
Open LLM & \nStrictDxii & \nSecondDxii & \nOutDxii & \precDstrictxii{} / \recDstrictxii & \precDxii{} / \recDxii & \precDbroadxii{} / \recDbroadxii \\
\midrule
\multicolumn{7}{l}{\emph{1921} (\nPagesFrame{} analysis pages)} \\
MLLM & \nStrictA & \nSecondA & \nOutA & \precAstrict{} / \recAstrict & \precA{} / \recA & \precAbroad{} / \recAbroad \\
Parser & \nStrictB & \nSecondB & \nOutB & \precBstrict{} / \recBstrict & \precB{} / \recB & \precBbroad{} / \recBbroad \\
Combined & \nStrictC & \nSecondC & \nOutC & \precCstrict{} / \recCstrict & \precC{} / \recC & \precCbroad{} / \recCbroad \\
Open LLM & \nStrictD & \nSecondD & \nOutD & \precDstrict{} / \recDstrict & \precD{} / \recD & \precDbroad{} / \recDbroad \\
\bottomrule
\end{tabular}
\begin{tablenotes}\footnotesize
\item \pdslabel{} Broad frame: 1912, \nPagesBroadxii{} page keys and \nRowsCleanedBroadxii{} cleaned rows; 1921, \nPagesBroad{} page keys and \nRowsCleanedBroad{} cleaned rows. Table~\ref{tab:records} reports the open OCR-LLM scores on the analysis frame. Source: the common scorer\textquotesingle s comparison results.
\end{tablenotes}
\end{threeparttable}
\end{table}

\begin{table}[!ht]\centering
\begin{threeparttable}
\caption{The MLLM workflow by stage, 1912 and 1921}\label{tab:stages}
\begin{tabular}{lrr}
\toprule
Stage & Precision (\%) & Recall (\%) \\
\midrule
\multicolumn{3}{l}{\emph{1912}} \\
Location pass alone (OCR text) & \precStageLocxii & \recStageLocxii \\
Biography pass alone (image) & \precStageBioxii & \recStageBioxii \\
Matched, before OCR enhancement & \precStageMatchNoRxii & \recStageMatchNoRxii \\
Matched, after OCR enhancement (MLLM workflow) & \precAxii & \recAxii \\
\midrule
\multicolumn{3}{l}{\emph{1921}} \\
Location pass alone (OCR text) & \precStageLoc & \recStageLoc \\
Biography pass alone (image) & \precStageBio & \recStageBio \\
Matched, before OCR enhancement & \precStageMatchNoR & \recStageMatchNoR \\
Matched, after OCR enhancement (MLLM workflow) & \precA & \recA \\
\bottomrule
\end{tabular}
\begin{tablenotes}\footnotesize
\item \pdslabel{} Analysis frames; second-pass links. The biography pass carries no geography and links on the first pass only. Source: the stage comparison.
\end{tablenotes}
\end{threeparttable}
\end{table}

\clearpage
\begin{landscape}
\begin{table}[!ht]\centering
\caption{Agreement by uncertainty flags, 1912 and 1921}\label{tab:strata}
\footnotesize
\setlength{\tabcolsep}{4pt}
\begin{tabular}{lrrrrrrr}
\toprule
Group & Reference rows & MLLM & Parser & Combined & Open LLM & Open OCR-LLM \\
\midrule
\multicolumn{7}{l}{\textbf{1912}} \\
\emph{Pages grouped by the share of entries flagged} (precision / recall) \\
$\geq$5\% of entries flagged (\nPagesUncHighxii{} pages) & \nRowsUncHighxii & \precUncHighAxii{} / \recUncHighAxii & \precUncHighBxii{} / \recUncHighBxii & \precUncHighCxii{} / \recUncHighCxii & \precUncHighDxii{} / \recUncHighDxii & \precUncHighExii{} / \recUncHighExii \\
$<$5\% of entries flagged (\nPagesUncLowxii{} pages) & \nRowsUncLowxii & \precUncLowAxii{} / \recUncLowAxii & \precUncLowBxii{} / \recUncLowBxii & \precUncLowCxii{} / \recUncLowCxii & \precUncLowDxii{} / \recUncLowDxii & \precUncLowExii{} / \recUncLowExii \\
\emph{Entries grouped by their own flag} (recall) \\
Flagged entries & \nRowsEntUncUncertainxii & \recEntUncUncertainAxii & \recEntUncUncertainBxii & \recEntUncUncertainCxii & \recEntUncUncertainDxii & \recEntUncUncertainExii \\
Unflagged entries & \nRowsEntUncCertainxii & \recEntUncCertainAxii & \recEntUncCertainBxii & \recEntUncCertainCxii & \recEntUncCertainDxii & \recEntUncCertainExii \\
\midrule
\multicolumn{7}{l}{\textbf{1921}} \\
\emph{Pages grouped by the share of entries flagged} (precision / recall) \\
$\geq$5\% of entries flagged (\nPagesUncHigh{} pages) & \nRowsUncHigh & \precUncHighA{} / \recUncHighA & \precUncHighB{} / \recUncHighB & \precUncHighC{} / \recUncHighC & \precUncHighD{} / \recUncHighD & \precUncHighE{} / \recUncHighE \\
$<$5\% of entries flagged (\nPagesUncLow{} pages) & \nRowsUncLow & \precUncLowA{} / \recUncLowA & \precUncLowB{} / \recUncLowB & \precUncLowC{} / \recUncLowC & \precUncLowD{} / \recUncLowD & \precUncLowE{} / \recUncLowE \\
\emph{Entries grouped by their own flag} (recall) \\
Flagged entries & \nRowsEntUncUncertain & \recEntUncUncertainA & \recEntUncUncertainB & \recEntUncUncertainC & \recEntUncUncertainD & \recEntUncUncertainE \\
Unflagged entries & \nRowsEntUncCertain & \recEntUncCertainA & \recEntUncCertainB & \recEntUncCertainC & \recEntUncCertainD & \recEntUncCertainE \\
\bottomrule
\end{tabular}
\par\smallskip
\begin{minipage}{\textwidth}\footnotesize
\emph{Notes:} \pdslabel{} Second-pass links on each edition's analysis frame (1912, \nPagesFramexii{} pages; 1921, \nPagesFrame{} pages), with the same links and scorer as Table~\ref{tab:records}. The first panel of each edition restricts both measures to the pages in its group; pages are grouped by the share of their reference entries that the transcribers flagged as uncertain. The second panel splits the reference entries by their own flag and reports recall only: an output row that matches no reference entry carries no flag, so precision cannot be assigned to an entry group. The flags record the transcribers' uncertainty; their absence does not verify an entry. The flagged entries include entries whose transcribed name contains a question mark. Matching uses the name, so no workflow can match these entries, and they count as unmatched for every workflow. They account for most of the low recall on flagged entries, especially in 1912, where about half of the flagged entries have such names.
\end{minipage}
\end{table}
\end{landscape}
\clearpage

\section{The MLLM workflow in detail}\label{app:gemini}

This appendix describes the stored 1921 implementation of the MLLM workflow (Section~\ref{sec:gemini}): the two model passes, the checks on their output and the matching that joins them.
Every call is to \geminiModelName{} through the vendor's API with a prompt written for the 1921 edition; the stored runs date from February 2026 (Appendix~\ref{app:repro}), and no call has been repeated since the model was shut down on \geminiShutdownDate.

\paragraph{Location pass.}
The first pass reads the OCR text of a page, not the image.
The prompt asks for structured JSON: the state, each town heading with its county and population, and the physician names printed beneath it.
A page whose output contains neither a town heading nor a physician name is skipped as presumed non-entry material; this rule can also exclude a physician page whose headings the pass fails to extract.

\paragraph{Biography pass.}
The second pass reads the page image.
The model first transcribes the page with its layout into sections, then extracts from each section the physicians and the fields the key defines.
A section that fails the schema check is sent back once for repair, and a section whose output overran the model's limit is split and read again.
Pages and entries with none of the four key fields (birth year, school, graduation year, license year) are dropped at collection.

\paragraph{Matching.}
The matcher joins the two passes in four greedy, one-to-one rounds: on the normalized name and its occurrence on the page; on the name alone by Jaro--Winkler similarity at or above 0.85; by carrying a location forward from the previous page to a biography that has none; and, last, by filling the fields of a name the location pass found but the image pass did not from the OCR text.
That last step is the only one at which OCR text supplies biographical values to this workflow; it filled \nRFourRowsxii{} rows in 1912 and \nRFourRows{} in 1921 (\nRFourRowsFramexii{} and \nRFourRowsFrame{} on the analysis frames).
Table~\ref{tab:stages} scores each stage on the analysis frames. In 1921, each pass alone links \rangeStageRec{} percent of the transcribed rows at about \rangeStagePrec{} percent precision, the two matched without the OCR step reach \recStageMatchNoR\% at \precStageMatchNoR\%, and the full extraction \recA\% at \precA\%.
The OCR step adds \nRFourLinksAddedxii{} links in 1912 and \nRFourLinksAdded{} in 1921 and about 1.4 points of recall in 1921, at a small cost in precision.

\section{Reproducibility statement}\label{app:repro}

The comparisons use stored outputs; no OCR call, model call or extraction was made for this note. Appendix~\ref{app:protocol} describes the linkage rule, scorer, frames and inputs. This appendix describes what is released and the limits to reproducing the results.

\paragraph{Extraction and release.}
The stored inputs include the OCR text, the MLLM workflow's biography and location passes and their matched output, the parser candidates and the combined tables. For 1921, OCR was performed in January 2026 and the model passes in February 2026, using \geminiModelName. For the combined table, county is taken from the separately checked county-cleaning rule on identical links; the other fields remain those of the reconciled output.
The public dataset is identified by Zenodo version \releaseVersion{} and DOI \zenodoDOI{}. The descriptive statistics (Table~\ref{tab:desc}) are calculated from the public release tables and can be reproduced from the public files.

\paragraph{Limits to reproduction.}
The model \geminiModelName{} was shut down on \geminiShutdownDate, so the MLLM extraction cannot be rerun through the vendor's API. The manual transcriptions are excluded from the public release, which prevents full public recomputation of the agreement results.

\section{Field tables by workflow}\label{app:fields}

Table~\ref{tab:fields} gives the percentages shown in Figure~\ref{fig:fields}, for both editions.

\begin{table}[!ht]\centering
\small
\begin{threeparttable}
\caption{Field agreement by workflow, 1912 and 1921 (precision / recall, \%)}\label{tab:fields}
\begin{tabular}{lrrrrr}
\toprule
Field & MLLM & Parser & Combined & Open LLM & Open OCR-LLM \\
\midrule
\multicolumn{6}{l}{\emph{Panel A: 1912}} \\
Name & \fieldNameAxii & \fieldNameBxii & \fieldNameCxii & \fieldNameDxii & \fieldNameExii \\
Birth year & \fieldBirthYearAxii & \fieldBirthYearBxii & \fieldBirthYearCxii & \fieldBirthYearDxii & \fieldBirthYearExii \\
Medical school & \fieldSchoolAxii & \fieldSchoolBxii & \fieldSchoolCxii & \fieldSchoolDxii & \fieldSchoolExii \\
Graduation year & \fieldGradYearAxii & \fieldGradYearBxii & \fieldGradYearCxii & \fieldGradYearDxii & \fieldGradYearExii \\
License year & \fieldLicenceYearAxii & \fieldLicenceYearBxii & \fieldLicenceYearCxii & \fieldLicenceYearDxii & \fieldLicenceYearExii \\
State & \fieldStateAxii & \fieldStateBxii & \fieldStateCxii & \fieldStateDxii & \fieldStateExii \\
City & \fieldCityAxii & \fieldCityBxii & \fieldCityCxii & \fieldCityDxii & \fieldCityExii \\
County & \fieldCountyAxii & \fieldCountyBxii & \fieldCountyCxii & \fieldCountyDxii & \fieldCountyExii \\
\addlinespace
\multicolumn{6}{l}{\emph{Panel B: 1921}} \\
Name & \fieldNameA & \fieldNameB & \fieldNameC & \fieldNameD & \fieldNameE \\
Birth year & \fieldBirthYearA & \fieldBirthYearB & \fieldBirthYearC & \fieldBirthYearD & \fieldBirthYearE \\
Medical school & \fieldSchoolA & \fieldSchoolB & \fieldSchoolC & \fieldSchoolD & \fieldSchoolE \\
Graduation year & \fieldGradYearA & \fieldGradYearB & \fieldGradYearC & \fieldGradYearD & \fieldGradYearE \\
License year & \fieldLicenceYearA & \fieldLicenceYearB & \fieldLicenceYearC & \fieldLicenceYearD & \fieldLicenceYearE \\
State & \fieldStateA & \fieldStateB & \fieldStateC & \fieldStateD & \fieldStateE \\
City & \fieldCityA & \fieldCityB & \fieldCityC & \fieldCityD & \fieldCityE \\
County & \fieldCountyA & \fieldCountyB & \fieldCountyC & \fieldCountyD & \fieldCountyE \\
\bottomrule
\end{tabular}
\begin{tablenotes}\footnotesize
\item \pdslabel{} Same frames, links and scorer as Table~\ref{tab:records}. Name is a link key, so its agreement is linkage-conditioned. Years as two-digit tokens; school codes after syntactic normalization; places after formatting normalization. Integer counts for both editions appear in Table~\ref{tab:fieldcounts}.
\end{tablenotes}
\end{threeparttable}
\end{table}

Table~\ref{tab:fieldcounts} gives the integer counts behind Table~\ref{tab:fields}: agreeing linked values, scorable output values and scorable reference values, for each field and the five workflows, on the 1912 and 1921 analysis frames.
Table~\ref{tab:directions} gives the disagreement counts of Section~\ref{sec:disagree}.
Table~\ref{tab:cleaning} shows what release cleaning does to the release table's agreement, scored as read and after cleaning on links fixed from the as-read values.
We selected the cleaning rules by comparing candidates against the transcription in the 1912 and 1921 editions before running this comparison.

\clearpage
\begin{landscape}
\begin{table}[!ht]\centering
\footnotesize
\caption{Field counts by workflow, 1912 and 1921: agreeing / scorable output / scorable reference values}\label{tab:fieldcounts}
\setlength{\tabcolsep}{2.5pt}
\begin{tabular}{lrrrrr}
\toprule
Field & MLLM & Parser & Combined & Open LLM & Open OCR-LLM \\
\midrule
\multicolumn{6}{l}{\emph{1912}} \\
Name & \cntNameAxii & \cntNameBxii & \cntNameCxii & \cntNameDxii & \cntNameExii \\
Birth year & \cntBirthYearAxii & \cntBirthYearBxii & \cntBirthYearCxii & \cntBirthYearDxii & \cntBirthYearExii \\
Medical school & \cntSchoolAxii & \cntSchoolBxii & \cntSchoolCxii & \cntSchoolDxii & \cntSchoolExii \\
Graduation year & \cntGradYearAxii & \cntGradYearBxii & \cntGradYearCxii & \cntGradYearDxii & \cntGradYearExii \\
License year & \cntLicenceYearAxii & \cntLicenceYearBxii & \cntLicenceYearCxii & \cntLicenceYearDxii & \cntLicenceYearExii \\
State & \cntStateAxii & \cntStateBxii & \cntStateCxii & \cntStateDxii & \cntStateExii \\
City & \cntCityAxii & \cntCityBxii & \cntCityCxii & \cntCityDxii & \cntCityExii \\
County & \cntCountyAxii & \cntCountyBxii & \cntCountyCxii & \cntCountyDxii & \cntCountyExii \\
\midrule
\multicolumn{6}{l}{\emph{1921}} \\
Name & \cntNameA & \cntNameB & \cntNameC & \cntNameD & \cntNameE \\
Birth year & \cntBirthYearA & \cntBirthYearB & \cntBirthYearC & \cntBirthYearD & \cntBirthYearE \\
Medical school & \cntSchoolA & \cntSchoolB & \cntSchoolC & \cntSchoolD & \cntSchoolE \\
Graduation year & \cntGradYearA & \cntGradYearB & \cntGradYearC & \cntGradYearD & \cntGradYearE \\
License year & \cntLicenceYearA & \cntLicenceYearB & \cntLicenceYearC & \cntLicenceYearD & \cntLicenceYearE \\
State & \cntStateA & \cntStateB & \cntStateC & \cntStateD & \cntStateE \\
City & \cntCityA & \cntCityB & \cntCityC & \cntCityD & \cntCityE \\
County & \cntCountyA & \cntCountyB & \cntCountyC & \cntCountyD & \cntCountyE \\
\bottomrule
\end{tabular}
\par\smallskip
\begin{minipage}{\textwidth}\footnotesize
\emph{Notes:} \pdslabel{} Unmatched rows' values stay in the denominators; a cell with several years or schools contributes each value. Source: the field-agreement comparison.
\end{minipage}
\end{table}
\end{landscape}
\clearpage

\begin{table}[!ht]\centering
\begin{threeparttable}
\caption{Release cleaning and agreement, combined table, 1912 and 1921 (precision / recall, \%)}\label{tab:cleaning}
\begin{tabular}{lrr}
\toprule
Field & As read & After cleaning \\
\midrule
\multicolumn{3}{l}{\emph{1912}} \\
Medical school & \cleanSchoolAsReadxii & \cleanSchoolCleanedxii \\
Birth year & \cleanBirthYearAsReadxii & \cleanBirthYearCleanedxii \\
City & \cleanCityAsReadxii & \cleanCityCleanedxii \\
County & \cleanCountyAsReadxii & \cleanCountyCleanedxii \\
\midrule
\multicolumn{3}{l}{\emph{1921}} \\
Medical school & \cleanSchoolAsRead & \cleanSchoolCleaned \\
Birth year & \cleanBirthYearAsRead & \cleanBirthYearCleaned \\
City & \cleanCityAsRead & \cleanCityCleaned \\
County & \cleanCountyAsRead & \cleanCountyCleaned \\
\bottomrule
\end{tabular}
\begin{tablenotes}\footnotesize
\item \pdslabel{} Links fixed from the as-read values. The as-read school, birth-year and city figures and the cleaned county figure are the combined table's entries in Table~\ref{tab:fields}; the comparison is between the release table's \texttt{*\_as\_read} columns and its cleaned columns. Source: the release-cleaning comparison.
\end{tablenotes}
\end{threeparttable}
\end{table}

\begin{table}[!ht]\centering
\small
\begin{threeparttable}
\caption{Disagreement between the MLLM workflow and the AI-assisted parser on rows linked through the combined table, 1912 and 1921}\label{tab:directions}
\begin{tabular}{lrrrr}
\toprule
Field & Both agree & Only MLLM agrees & Only parser agrees & Neither agrees \\
\midrule
\multicolumn{5}{l}{\emph{1912}} \\
Birth year & \disBirthYearBothAgreexii & \disBirthYearGeminiRightxii & \disBirthYearParserRightxii & \disBirthYearNeitherxii \\
Graduation year & \disGradYearBothAgreexii & \disGradYearGeminiRightxii & \disGradYearParserRightxii & \disGradYearNeitherxii \\
License year & \disLicenceYearBothAgreexii & \disLicenceYearGeminiRightxii & \disLicenceYearParserRightxii & \disLicenceYearNeitherxii \\
Medical school & \disSchoolBothAgreexii & \disSchoolGeminiRightxii & \disSchoolParserRightxii & \disSchoolNeitherxii \\
\midrule
\multicolumn{5}{l}{\emph{1921}} \\
Birth year & \disBirthYearBothAgree & \disBirthYearGeminiRight & \disBirthYearParserRight & \disBirthYearNeither \\
Graduation year & \disGradYearBothAgree & \disGradYearGeminiRight & \disGradYearParserRight & \disGradYearNeither \\
License year & \disLicenceYearBothAgree & \disLicenceYearGeminiRight & \disLicenceYearParserRight & \disLicenceYearNeither \\
Medical school & \disSchoolBothAgree & \disSchoolGeminiRight & \disSchoolParserRight & \disSchoolNeither \\
\bottomrule
\end{tabular}
\begin{tablenotes}\footnotesize
\item \pdslabel{} Cells on the combined table's linked rows: 1912, \nLinkedRowsVCxii{} rows on \nLinkedPagesVCxii{} pages; 1921, \nLinkedRowsVC{} rows on \nLinkedPagesVC{} pages. An absent extraction entry counts as non-agreement. Counts describe agreement with the transcription and do not establish which extraction reproduces the printed entry. Rows where the reference is blank are left out; a cell with several school codes counts once here and once per value in Table~\ref{tab:fieldcounts}. Source: the comparison of extraction disagreements.
\end{tablenotes}
\end{threeparttable}
\end{table}

\section{Cost accounting}\label{app:cost}

Table~\ref{tab:costs} gives the cost of manual transcription and the direct processing charges for the machine workflows, in total and per page.
Manual costs are the delivered workbook page counts, \costManualPagesBoth{} pages across both editions, at the invoiced \texteuro{}\costPdsPageEur{} per page. The machine cost per page divides the combined Gemini and Azure charges by the same page count. We allocate Gemini charges by whole billing days and assume one successful Azure OCR pass per page, so the machine figures are estimates, not invoice totals. Ratios use unrounded inputs and one exchange-rate date (ECB, 3 February 2026).

The comparison covers direct processing charges only. It leaves out engineering, review, model development, coding-agent assistance, energy and the local computing hardware. The hardware is shared with other project work and hosted-API workflows do not need it, so adding it would not make the comparison like for like. The manual transcription also covers a different set of pages and fields than the machine outputs, and we do not estimate a separate cost for the open model pipelines.

\begin{table}[!ht]\centering
\small
\begin{threeparttable}
\caption{Manual transcription and estimated machine processing charges}\label{tab:costs}
\begin{tabular}{llrrr}
\toprule
Component & Unit & 1912 & 1921 & Both editions \\
\midrule
Purchased transcription & Pages & \costManualPagesxii & \costManualPages & \costManualPagesBoth \\
Manual transcription & EUR & \costManualEurxii & \costManualEur & \costManualEurBoth \\
Gemini, allocated billing & DKK & \costGeminiDkkxii & \costGeminiDkk & \costGeminiDkkBoth \\
Azure, one-pass estimate & USD & \costAzureUsdxii & \costAzureUsd & \costAzureUsdBoth \\
Gemini plus Azure & EUR & \costApiEurxii & \costApiEur & \costApiEurBoth \\
Manual cost per page & EUR & \costPdsPageEur & \costPdsPageEur & \costPdsPageEur \\
Machine cost per page & EUR & \costApiPageEurxii & \costApiPageEur & \costApiPageEurBoth \\
Manual / estimated API cost & Ratio & \costApiRatioxii & \costApiRatio & \costApiRatioBoth \\
\bottomrule
\end{tabular}
\begin{tablenotes}\footnotesize
\item Manual costs use delivered workbook page counts at \texteuro{}\costPdsPageEur{} per page, reconciled to the vendor's invoices; the 1918 batch and unassigned supplement are excluded. These are purchased-coverage counts, not the analysis frames. API figures cover Gemini and Azure, not a separate cost estimate for the open model pipelines.
\item Gemini 2.0 Flash billing is allocated by whole days: 28--31 January 2026 (1912), 23--25 January and 3--5 February (1921); other-model charges are excluded. Edition overlap on 25 and 31 January makes attribution approximate. Excluding 31 January gives DKK\,\costGeminiSharedExcludedDkkxii{} for 1912; excluding 25 January gives DKK\,\costGeminiSharedExcludedDkk{} for 1921. These are scenarios, not confidence bounds.
\item Azure assumes \costOcrPagesxii{} and \costOcrPages{} successful one-page images at the conditional West Europe prebuilt-layout S0 list price of USD\,\costAzureThousandUsd{} per 1,000 pages (effective from August 2021). Actual Azure billing and repeat-call counts were not available; these are list-price estimates, not invoice totals. EUR equivalents and ratios use ECB reference rates on 3 February 2026: EUR\,1 = DKK\,\costFxDkkEur{} = USD\,\costFxUsdEur{}, a common-date presentation convention; totals and ratios use unrounded inputs.
\end{tablenotes}
\end{threeparttable}
\end{table}
\FloatBarrier

\section{Robustness to the page frame}\label{app:robust}

The analysis frames keep every transcribed page without an ignored-row placeholder (Section~\ref{sec:reference}). Two choices could move the results: leaving out the pages flagged for incomplete coverage, and keeping the pages seen during parser development, drawn into the frozen design sample or used in diagnostics. Table~\ref{tab:robust-coverage} scores every workflow on all transcribed pages and on the incomplete pages alone; Table~\ref{tab:robust-exposure} scores them on the analysis frame without the exposed pages and on the exposed pages alone. Links are fixed edition-wide before any page restriction, so each frame scores the same links.

\begin{table}[!ht]\centering
\small
\begin{threeparttable}
\caption{Record agreement by coverage frame, 1912 and 1921}\label{tab:robust-coverage}
\footnotesize
\setlength{\tabcolsep}{3pt}
\begin{tabular}{lrrrrrrrr}
\toprule
Coverage & \multicolumn{4}{c}{1912} & \multicolumn{4}{c}{1921} \\
\cmidrule(lr){2-5}\cmidrule(lr){6-9}
 & \multicolumn{2}{c}{All transcribed} & \multicolumn{2}{c}{Incomplete pages} & \multicolumn{2}{c}{All transcribed} & \multicolumn{2}{c}{Incomplete pages} \\
\cmidrule(lr){2-3}\cmidrule(lr){4-5}\cmidrule(lr){6-7}\cmidrule(lr){8-9}
Workflow & Prec.\ (\%) & Recall (\%) & Prec.\ (\%) & Recall (\%) & Prec.\ (\%) & Recall (\%) & Prec.\ (\%) & Recall (\%) \\
\midrule
MLLM & \precRobAllAxii & \recRobAllAxii & \precRobIncompleteAxii & \recRobIncompleteAxii & \precRobAllA & \recRobAllA & \precRobIncompleteA & \recRobIncompleteA \\
Parser & \precRobAllBxii & \recRobAllBxii & \precRobIncompleteBxii & \recRobIncompleteBxii & \precRobAllB & \recRobAllB & \precRobIncompleteB & \recRobIncompleteB \\
Combined & \precRobAllCxii & \recRobAllCxii & \precRobIncompleteCxii & \recRobIncompleteCxii & \precRobAllC & \recRobAllC & \precRobIncompleteC & \recRobIncompleteC \\
Open LLM & \precRobAllDxii & \recRobAllDxii & \precRobIncompleteDxii & \recRobIncompleteDxii & \precRobAllD & \recRobAllD & \precRobIncompleteD & \recRobIncompleteD \\
Open OCR-LLM & \precRobAllExii & \recRobAllExii & \precRobIncompleteExii & \recRobIncompleteExii & \precRobAllE & \recRobAllE & \precRobIncompleteE & \recRobIncompleteE \\
\bottomrule
\end{tabular}
\begin{tablenotes}\footnotesize
\item \pdslabel{} All transcribed: \nPagesRobAll{} pages and \nRowsRobAll{} reference rows in 1921, \nPagesRobAllxii{} and \nRowsRobAllxii{} in 1912. Incomplete pages: the pages with an ignored-row placeholder, \nPagesRobIncomplete{} pages and \nRowsRobIncomplete{} rows in 1921, \nPagesRobIncompletexii{} and \nRowsRobIncompletexii{} in 1912. On these pages part of the printed entries were not transcribed, so precision understates agreement. Links after the second pass; open LLM and open OCR-LLM as in Table~\ref{tab:records}.
\end{tablenotes}
\end{threeparttable}
\end{table}

\begin{table}[!ht]\centering
\small
\begin{threeparttable}
\caption{Record agreement with and without exposed pages, 1912 and 1921}\label{tab:robust-exposure}
\footnotesize
\setlength{\tabcolsep}{3pt}
\begin{tabular}{lrrrrrrrr}
\toprule
Exposure & \multicolumn{4}{c}{1912} & \multicolumn{4}{c}{1921} \\
\cmidrule(lr){2-5}\cmidrule(lr){6-9}
 & \multicolumn{2}{c}{Without exposed pages} & \multicolumn{2}{c}{Exposed pages only} & \multicolumn{2}{c}{Without exposed pages} & \multicolumn{2}{c}{Exposed pages only} \\
\cmidrule(lr){2-3}\cmidrule(lr){4-5}\cmidrule(lr){6-7}\cmidrule(lr){8-9}
Workflow & Prec.\ (\%) & Recall (\%) & Prec.\ (\%) & Recall (\%) & Prec.\ (\%) & Recall (\%) & Prec.\ (\%) & Recall (\%) \\
\midrule
MLLM & \precRobOldAxii & \recRobOldAxii & \precRobExposedAxii & \recRobExposedAxii & \precRobOldA & \recRobOldA & \precRobExposedA & \recRobExposedA \\
Parser & \precRobOldBxii & \recRobOldBxii & \precRobExposedBxii & \recRobExposedBxii & \precRobOldB & \recRobOldB & \precRobExposedB & \recRobExposedB \\
Combined & \precRobOldCxii & \recRobOldCxii & \precRobExposedCxii & \recRobExposedCxii & \precRobOldC & \recRobOldC & \precRobExposedC & \recRobExposedC \\
Open LLM & \precRobOldDxii & \recRobOldDxii & \precRobExposedDxii & \recRobExposedDxii & \precRobOldD & \recRobOldD & \precRobExposedD & \recRobExposedD \\
Open OCR-LLM & \precRobOldExii & \recRobOldExii & \precRobExposedExii & \recRobExposedExii & \precRobOldE & \recRobOldE & \precRobExposedE & \recRobExposedE \\
\bottomrule
\end{tabular}
\begin{tablenotes}\footnotesize
\item \pdslabel{} Without exposed pages: the analysis frame less every page with a scan seen while tuning the parser, drawn into the frozen design sample or used in diagnostics, \nPagesRobOld{} pages and \nRowsRobOld{} reference rows in 1921, \nPagesRobOldxii{} and \nRowsRobOldxii{} in 1912. Exposed pages only: the \nPagesRobExposed{} and \nPagesRobExposedxii{} pages left out there (\nRowsRobExposed{} and \nRowsRobExposedxii{} rows): in 1921, \nPagesExposedDesign{} design, \nPagesExposedTuning{} tuning and \nPagesExposedDiagnostic{} diagnostic pages; in 1912, \nPagesExposedDesignxii{}, \nPagesExposedTuningxii{} and \nPagesExposedDiagnosticxii{}. A page on several lists counts once, as design before tuning before diagnostic. The record of pages seen during parser development is incomplete for 1921. Links after the second pass.
\end{tablenotes}
\end{threeparttable}
\end{table}

\FloatBarrier
\end{document}